\documentclass[aps,prd,twocolumn,amsmath,superscriptaddress,amssymb,showpacs,floatfix,nofootinbib,longbibliography]{revtex4-1}
\pdfoutput=1
\usepackage{graphicx}
\usepackage{bm}
\usepackage{times}
\usepackage[dvipsnames]{xcolor}
\usepackage{hyperref}
\hypersetup{
    colorlinks = true,
    linkcolor = {MidnightBlue}, 
    citecolor = {MidnightBlue},
    urlcolor = {MidnightBlue},
}
\usepackage{slashed}
\usepackage{color}
\usepackage{aas_macros}
\usepackage{slashed}
\usepackage{lipsum}
\usepackage{subfigure}
\usepackage{multirow}
\usepackage{amsmath}
\usepackage{array} % for defining a new column type
\usepackage{varwidth} %for the varwidth minipage environment

\newcommand{\be}{\begin{equation}}
\newcommand{\ee}{\end{equation}}
\newcommand{\bea}{\begin{eqnarray}}
\newcommand{\eea}{\end{eqnarray}}

\begin{document}

\title{Accurate Inverse-Compton Models Strongly Enhance Leptophilic Dark Matter Signals}
\author{Isabelle John}
\email{isabelle.john@fysik.su.se, ORCID: orcid.org/0000-0003-2550-7038}
\affiliation{Stockholm University and The Oskar Klein Centre for Cosmoparticle Physics,  Alba Nova, 10691 Stockholm, Sweden}
\author{Tim Linden}
\email{linden@fysik.su.se, ORCID: orcid.org/0000-0001-9888-0971}
\affiliation{Stockholm University and The Oskar Klein Centre for Cosmoparticle Physics,  Alba Nova, 10691 Stockholm, Sweden}

\begin{abstract}
The annihilation of TeV-scale leptophilic dark matter into electron-positron pairs (hereafter $e^+e^-$) will produce a sharp cutoff in the local cosmic-ray $e^+e^-$ spectrum at an energy matching the dark matter mass. At these high energies, $e^+e^-$ cool quickly due to synchrotron interactions with magnetic fields and inverse-Compton scattering with the interstellar radiation field. These energy losses are typically modelled as a continuous process. However, inverse-Compton scattering is a stochastic energy-loss process where interactions are rare but catastrophic. We show that when inverse-Compton scattering is modelled as a stochastic process, the expected $e^+e^-$ flux from dark matter annihilation is about a factor of $\sim$2 larger near the dark matter mass than in the continuous model. This greatly enhances the detectability of heavy dark matter annihilating to $e^+e^-$ final states.
\end{abstract}

\maketitle

\section{Introduction}
Dark matter particles that annihilate into electrons and positron pairs (hereafter, $e^+e^-$) are expected to produce a sharp spectral cutoff in the local cosmic-ray $e^+e^-$ spectrum at an energy that corresponds to the dark matter mass~\cite{Turner:1989kg, HESS:2008ibn, Bergstrom:2008gr, Cholis:2008qq, Cirelli:2008pk, Arkani-Hamed:2008hhe, Ibarra:2008jk, Profumo:2009uf, Dugger:2010ys, Bergstrom:2013jra, John:2021ugy, Coogan:2019uij}. While no such signal has been conclusively observed, the cosmic-ray $e^+e^-$ fluxes have been measured to great precision at energies up to $\sim$1~TeV (e.g. AMS-02~\cite{AMS:2019iwo, AMS:2019rhg}, H.E.S.S.~\cite{HESS:2008ibn, HESSdata}) and upcoming experiments are expected to reach energies of several tens to hundreds of TeV, e.g. Cherenkov Telescope Array (CTA)~\cite{CTAwebsite, Knodlseder:2020onx, 2013APh....43..171B, maier2019performance}, AMS-100~\cite{Schael:2019lvx} and HERD~\cite{HERD:2014bpk}, expanding the parameter space of the search for dark matter signals to even higher energies with unprecedented precision.

Observations indicate that there are several important components of the local $e^+e^-$ flux. At GeV energies, the dominant $e^+e^-$ comes from the secondary interactions of other cosmic rays. At higher energies between $\sim 20$~GeV to $\sim~400$~GeV an unexpected excess was measured in the positron flux by PAMELA~\cite{PAMELA:2013vxg} and AMS-02~\cite{PhysRevLett.122.041102}. While dark matter models of this excess have long-been explored~\cite{Turner:1989kg, HESS:2008ibn, Cirelli:2008pk, Arkani-Hamed:2008hhe, Profumo:2009uf, Dugger:2010ys}, current studies indicate that this excess is best explained by nearby pulsars that produce $e^+e^-$ pairs as they spin down (e.g.~\cite{Yuksel:2008rf, Hooper:2008kg, Profumo:2008ms, HAWC:2017kbo, Fang:2018qco, Hooper:2017gtd}). However, dark matter annihilation may produce a sub-dominant portion of the signal~\cite{Bergstrom:2013jra, John:2021ugy, Coogan:2019uij}. 

\begin{figure}[tb]
    \centering
    \includegraphics[width=0.48\textwidth]{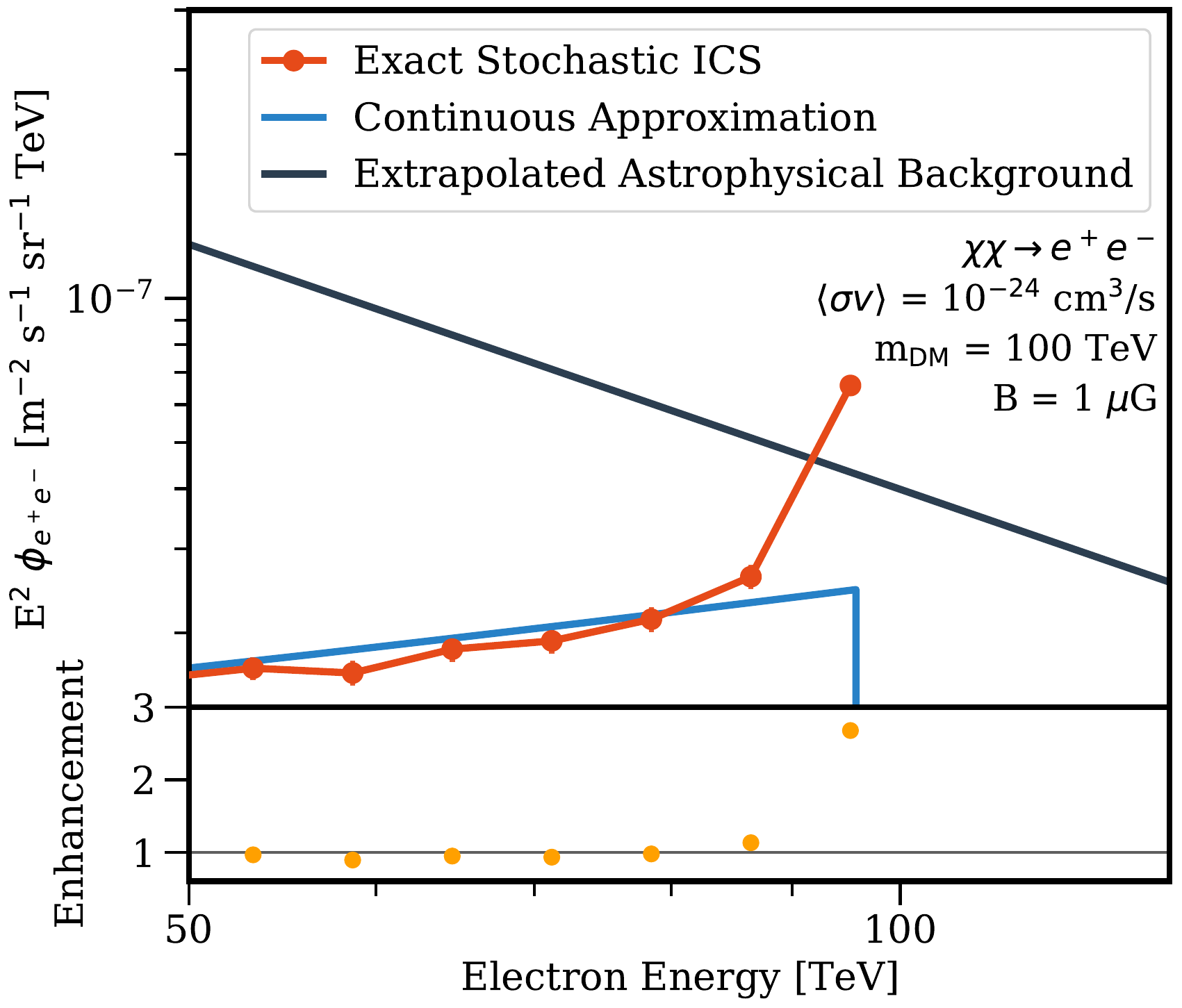}
    \caption{The local $e^+e^-$ flux expected from an annihilating dark matter particle with a mass of 100 TeV, for an energy loss model that approximates energy losses as either a continuous process (blue), or uses an exact stochastic formalism (red). The $e^+e^-$ flux sharpens at the energy near the dark matter mass by about a factor of 2.6 in the stochastic model compared to the continuous approximation. For reference, an astrophysical background is given by extrapolating H.E.S.S. data~\cite{HESSdata} to higher energies.}
    \label{fig: fig1}
\end{figure}

The propagation of the $e^+e^-$ produced in the dark matter annihilation event must be properly modelled to predict the observed $e^+e^-$ flux at Earth. During propagation, the $e^+e^-$ lose energy through several processes. The most relevant are synchrotron interactions with Galactic magnetic fields and inverse-Compton scattering (ICS) interactions with interstellar radiation fields (ISRF). Typically, these energy losses have been calculated under the assumption that they are continuous over time. This approach is approximately correct for synchrotron losses. However, ICS interactions are rare and can remove a large fraction of the $e^+e^-$ energy in just a single interaction. 

In a previous paper~\cite{John:2022asa}, we have investigated the impact of correctly taking into account the stochastic effects of ICS energy losses. Specifically, we have looked at the expected contribution from nearby pulsars to the positron flux, for which the continuous energy loss models predict sharp spectral features corresponding to the age of each pulsar. However, correctly treating the stochastic nature of ICS losses smooths out the local $e^+e^-$ spectrum. This can be understood when looking at $e^+e^-$ that are injected at the same initial energy and cool for the same amount of time: In the stochastic model, $e^+e^-$ cool to a distribution of final energies, while in the continuous approximation the initial $e^+e^-$ energy and pulsar age exactly determine the observed final energy.

Importantly, the stochasticity of ICS does not only affect the spectrum of $e^+e^-$ that originate from pulsars. It is relevant for any source that injects very-high-energy $e^+e^-$. One particularly interesting scenario involves TeV-scale leptophilic dark matter candidates, which has been extensively studied as potential solutions to the PAMELA positron excess~\cite{Arkani-Hamed:2008hhe}, the DAMPE 1.4~TeV excess~\cite{DAMPE:2017fbg, Duan:2017pkq}, as well as theoretically motivated models which have implications for, e.g., neutrino masses, the anomalous muon dipole moment, or well-motivated dark sector extensions of the standard model~\cite{Hambye:2006zn, Dev:2013hka, Agrawal:2014ufa, Bell:2014tta, Chakraborti:2020zxt}. These scenarios also generically predict a sharp spectral feature in the observed $e^+e^-$ flux at an energy that corresponds to the dark matter mass. Although the dark matter origin of the PAMELA and DAMPE $e^+e^-$ excesses are now disfavored by other observations, e.g.~\cite{Barger:2009yt, Malyshev:2009tw, Huang:2017egk}, this predicted distinct feature makes the local cosmic-ray $e^+e^-$ fluxes a powerful tool to search for indirect dark matter signals.

In this paper, we show that the correct stochastic treatment of ICS strongly enhances the spectral peak observed in the local $e^+e^-$ flux at an energy corresponding to the dark matter mass. This is due to the fact that in the stochastic case, $e^+e^-$ remain at their injected energy for a long time before they undergo their first interaction, which then instantaneously removes a large fraction of their total energy. By contrast, the continuous case smears out these energy losses across all $e^+e^-$, smoothing out the spectrum near the dark matter mass. The amplitude of this effect sensitively depends on the magnetic field strength, which produces continuous synchrotron losses in each case. In Figure~\ref{fig: fig1}, we show this effect for a dark matter mass of 100~TeV and a magnetic field strength of 1~$\mu$G, finding an enhancement in the sharpness of the local $e^+e^-$ spectrum by approximately a factor of 2.6.

This paper is structured as follows: In Section~\ref{sec: methodology}, we give a detailed background about the energy loss processes and describe the continuous~(\ref{sec: continuous model}) and stochastic~(\ref{sec: stochastic model}) energy loss models. In Section~\ref{sec: results}, we present our results for different dark matter masses, final states and magnetic field strengths. We summarise and discuss our results in Section~\ref{sec: discussion}.

\section{Methodology}
\label{sec: methodology}
In this section, we compute the relevant energy loss processes and describe our models for the calculation of the continuous ICS energy losses (Section~\ref{sec: continuous model}) and the stochastic ICS energy losses (Section~\ref{sec: stochastic model}).

In both cases, synchrotron energy loss rate is given by: 

\begin{equation}\label{eq: synchrotron losses}
\frac{dE_e}{dt} = \frac{4}{3} \sigma_T c \left(\frac{E_e}{m_e}\right) u_B,
\end{equation}
where $u_B$ is the energy density of the magnetic field, obtained from the magnetic field strength (in units of G) through ${ u_B = B^2 / (8\pi) \times 6.24 \times 10^{11} }$~eV/cm$^3$. The average energy loss in a synchrotron interaction for an $e^+e^-$ of energy $E_e$ is given by the critical energy

\begin{equation}\label{eq: synchrotron critical energy}
E_\text{crit, sync} = \frac{3\gamma^2eB}{4\pi m_e c} \approx 0.06 \left(\frac{B}{1\,\mu\text{G}}\right) \left(\frac{E_e}{1\text{ TeV}}\right)^2 \text{ eV} 
\end{equation}
which is $\sim 600$~eV for a magnetic field strength of ${ B = 1\,\mu }$G and an electron energy of ${ E_e = 100 }$~TeV. Even for strong magnetic fields of ${ B = 3~\mu}$G and ${ E_e = 300 }$~TeV, this is 16~keV, which is very small compared to the instrumental energy resolution. However, synchrotron interactions happen frequently and depending on the exact magnetic field strengths and ISRF components, synchrotron cooling typically exceeds ICS at energies exceeding $\sim 100$~TeV, when ICS becomes highly suppressed by Klein-Nishina effects.

For ICS processes, on the other hand, a high-energy $e^+e^-$ interacts with a photon from the interstellar radiation field, mostly from lower energy components such as the cosmic microwave background (CMB) or infrared (IR) radiation. The interaction cross section is given by~\cite{KN:1928, Blumenthal:1970gc, 1981Ap&SS..79..321A}:

\begin{multline}
    \label{eq:fullkn}
    \frac{d^2\sigma(E_\gamma, \theta)}{d\Omega dE_\gamma} = \frac{r_0^2}{2\nu_iE_e^2} ~ \times ~ \\ \left [1 +   \frac{z^2}{2(1-z)} - \frac{2z}{b_\theta(1-z)} +  \frac{2z^2}{b_\theta^2(1-z)^2}\right ]
\end{multline}
where $E_\gamma$ is energy of the outgoing $\gamma$-ray photon, $\nu_i$ is the initial energy of the photon, $E_e$ the initial energy of the $e^+e^-$, $\theta$ the scattering angle, $r_0$ the classical electron radius, ${ z \equiv E_\gamma/E_e }$ and ${ b_\theta \equiv 2 (1-\cos \theta) \nu_iE_e }$. This corresponds to a total energy loss rate for an $e^+e^-$ that is given by~\cite{Schlickeiser:2009qq}

\begin{equation}\label{eq: full KN}
\frac{dE_e}{dt} = \frac{12 c \sigma_T}{m_e^2} E_e^2 \int_0^\infty \nu n\left(\nu\right) J\left(\Gamma\right) \, d\nu,
\end{equation}
where $\sigma_T$ is the Thomson cross section, $\gamma = E_e/m_e$, $n(\nu)$ is the energy spectrum of the ISRF photons, and $J\left(\Gamma\right)$ corresponds to the suppression of the Thomson cross-section due to Klein-Nishina effects and is given by:

\begin{equation}
J\left(\Gamma\right) = \int_0^1 \frac{q G\left(q, \Gamma\right)}{\left(1 + \Gamma q\right)^3},
\end{equation}
where ${ \Gamma = 4\nu\gamma/m_e }$ and ${ q = \nu_s/(\Gamma(\gamma m - \nu_s)) }$, where $\nu_s$ is the energy of the scattered $\gamma$-ray photon. The function $G(q, \Gamma)$ is given by:

\begin{equation}
G\left(q, \Gamma\right) = 2q\ln{q} + (1+2q)(1-q) + \frac{\Gamma^2 q^2 (1-q)}{2\left(1+\Gamma q\right)}.
\end{equation}

From these equations it can be seen that ICS processes are energy dependent, \textit{i.e.}, the interaction cross section decreases at high energies (Klein-Nishina suppression), which makes interactions especially rare, while at the same time a single interaction takes an increasingly large fraction of the $e^+e^-$ energy. This means that ICS is a highly stochastic process, while synchrotron losses can be well-described as a continuous process.

\begin{figure}[tbp]
\centering
\includegraphics[width=0.48\textwidth]{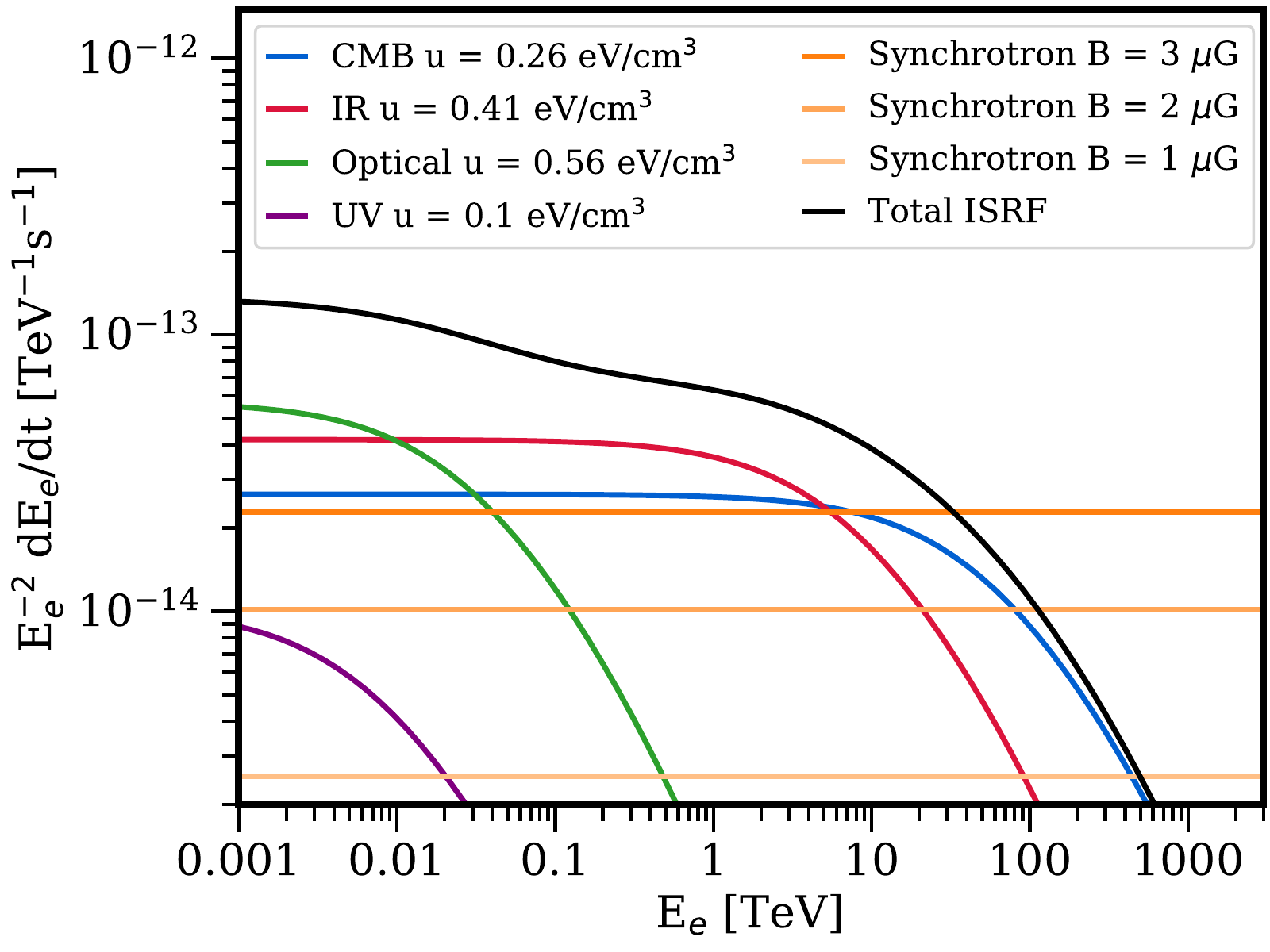}
\caption{The energy loss factor as a function of electron energy for the specific ISRF components, as well as the three magnetic field strengths used throughout this work. The black line shows the total energy losses from all ISRF components combined (\textit{i.e.}, CMB, infrared, optical and ultraviolet). It can be seen that synchrotron losses start to dominate over ICS losses for energies above a few hundred TeV, depending on the magnetic field strength.}
\label{fig: ISRF}
\end{figure}

Throughout this work, we use an interstellar radiation field similar to the model employed in the \texttt{Galprop} cosmic-ray propagation code~\cite{Porter:2008ve}, with four components: CMB, infrared (IR), optical/starlight and ultraviolet (UV) radiation. We assume the following energy densities and temperatures for each component:
${ u_\text{UV} = 0.1 }$~eV/cm$^3$, ${ T_\text{UV} = 20\,000 }$~K,
${ u_\text{optical} = 0.56 }$~eV/cm$^3$, ${ T_\text{optical} = 5\,000 }$~K,
${ u_\text{IR} = 0.41 }$~eV/cm$^3$, ${ T_\text{IR} = 20 }$~K,
${ u_\text{CMB} = 0.26 }$~eV/cm$^3$, ${ T_\text{CMB} = 2.7 }$~K. We note that for most of the dark masses considered here, only the CMB and IR energy densities have any effect on our results. Figure~\ref{fig: ISRF} shows the energy loss rates for the different ISRF components, as well as the energy losses for the three different magnetic field strengths (1~$\mu$G, 2~$\mu$G and 3~$\mu$G) that we consider throughout this paper.

\subsection{Continuous ICS Energy Loss Model}
\label{sec: continuous model}
In standard approaches, ICS energy losses are assumed to be continuous, taking an infinitesimal amount of energy from the $e^+e^-$ in infinitesimal time steps according to the energy loss rates given in Equations~\ref{eq: synchrotron losses} and~\ref{eq: full KN} for synchrotron and ICS losses, respectively.

We model this by taking an $e^+e^-$ with some initial energy and applying Equations~\ref{eq: synchrotron losses} and~\ref{eq: full KN} repeatedly for appropriately small times steps, until the desired cooling time has passed.

\subsection{Stochastic ICS Energy Loss Model}
\label{sec: stochastic model}
In our stochastic model, we treat ICS interactions precisely as a probabilistic process, rather than approximating it as a continuous process. For this we create a Monte Carlo setup to determine if an $e^+e^-$ undergoes an ICS interaction in a certain period of time, what the energy of the corresponding ISRF photon is, and how much energy is transferred in the interaction.

Specifically, we simulate the energy-loss evolution for each $e^+e^-$ individually by applying the following steps, as also discussed in~\cite{John:2022asa}: First, an $e^+e^-$ is injected at some initial energy. Then we calculate a time step to be sufficiently small so that the energy loss due to synchrotron processes and the probability of having two ICS interactions is negligible within that time step. Based on the Klein-Nishina cross section (Equation~\ref{eq: full KN}), we use a Monte Carlo to determine if an interaction happens in the time step, and, if an interaction happens, the energy of the ISRF photon. Then, using another Monte Carlo, we determine the magnitude of the energy loss in that interaction. Finally, we subtract the energy losses from any ICS that happens as well as a continuous energy loss from synchrotron radiation (Equation~\ref{eq: synchrotron losses}) during that time step. We repeat this process until the $e^+e^-$ have cooled for the desired amount of time.

\begin{figure*}[tbp]
\begin{minipage}[t]{0.48\textwidth}
\centering
\includegraphics[width=0.98\textwidth]{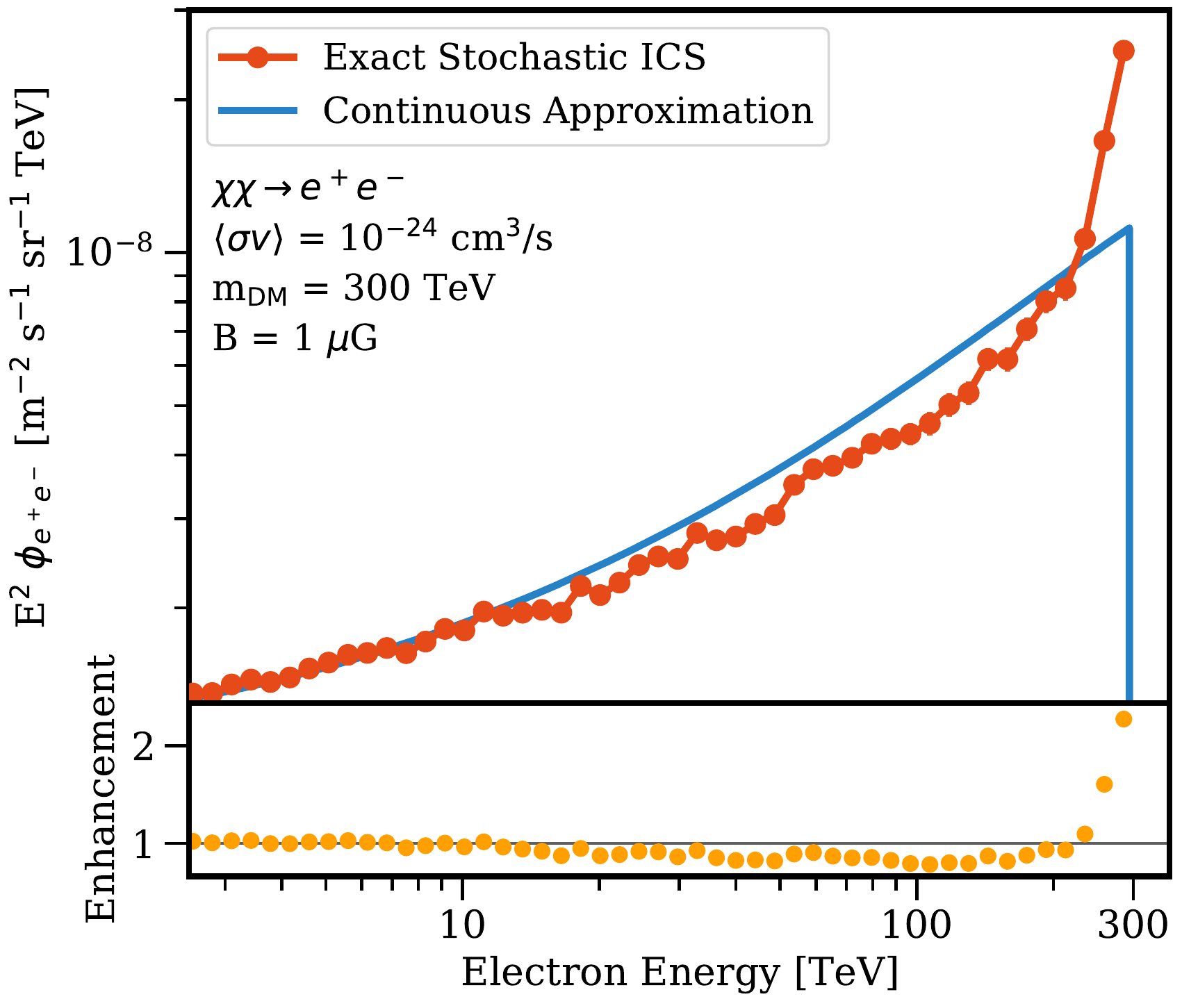}
\caption{The $e^+e^-$ flux expected from an annihilating dark matter particle with a mass of 300~TeV for a magnetic field strength of 1~$\mu$G. The result of the stochastic energy loss model is given in red, while the result of the continuous approximation model is given in blue. In the bottom, the enhancement of the sharp cutoff near the dark matter mass is given, showing that the feature is enhanced by about a factor of 2.3 in the stochastic model compared to the continuous treatment.}
\label{fig: 300 TeV 1 muG}
\end{minipage}
\hfill
\begin{minipage}[t]{0.48\textwidth}
\centering
\includegraphics[width=0.98\textwidth]{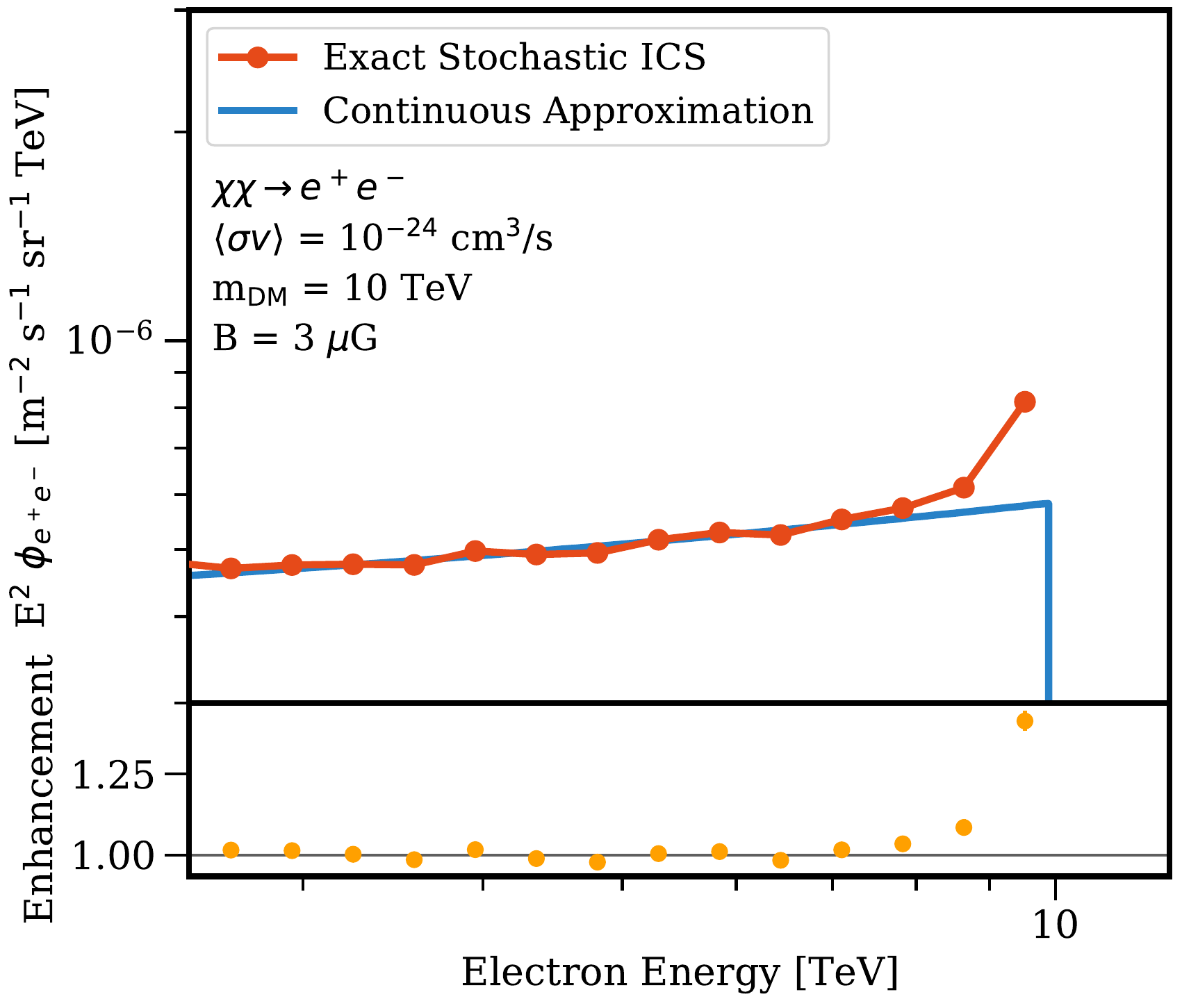}
\caption{The $e^+e^-$ flux expected from an annihilating dark matter particle with a mass of 10~TeV for a magnetic field strength of 3~$\mu$G. The result of the stochastic energy loss model is given in red, while the result of the continuous approximation model is given in blue. In the bottom, the enhancement of the sharp cutoff near the dark matter mass is given, showing that the feature is enhanced by about a factor of 1.4 in the stochastic model compared to the continuous treatment.}
\label{fig: 10 TeV 3 muG}
\end{minipage}
\end{figure*}

\subsection{Dark Matter Modeling}
\label{sec: injection spectra}
To obtain the flux produced by dark matter annihilation, we inject $e^+e^-$ for uniformly distributed random cooling times, with a maximum value that exceeds the $e^+e^-$ cooling times, in order to simulate continuous injection.

In this work, we consider two dark matter annihilation final states. In our main study, we assume that the dark matter particles annihilate directly into an $e^+e^-$ pair. This means that all $e^+e^-$ are injected at a single energy corresponding to the dark matter mass.

We further investigate a case where the dark matter particles annihilate into a $\mu^+\mu^-$ pair, that subsequently produces $e^+e^-$. In this case, the $e^+e^-$ are injected with a distribution of initial energies. To obtain this distribution, we use the injection spectra provided by
DarkSUSY~\cite{Bringmann:2018lay, Gondolo:2004sc, darksusywebsite}. Since DarkSUSY only includes these spectra up to dark matter masses of a few TeV, we re-scale the injection spectra to match the heavier dark matter masses that are of interest here. This is possible because at these high energies, the muon mass ($\sim$~106~MeV) is negligible.

\subsection{Electron and Positron Fluxes at Earth}
After obtaining the $e^+e^-$ fluxes from the simulations, we normalize them according to the dark matter annihilation rate. The rate of $e^+e^-$ production from annihilating dark matter particles is given by

\begin{equation}\label{eq: positron rate}
\frac{dn_e}{dt} = \frac{1}{2} \left(\frac{\rho_0}{m_\text{DM}}\right)^2 \langle\sigma v\rangle \; \frac{dN_e}{dE_e},
\end{equation}
where $dn_e/dt$ is the number density of $e^+e^-$ per unit time, per unit volume, per energy, $\rho_0$ is the local dark matter energy density, $m_\text{DM}$ the mass of the dark matter particle, $\langle \sigma v\rangle$ the thermally averaged dark matter annihilation cross section, and $dN_e/dE_e$ is the energy spectrum of $e^+e^-$ (differential number density of $e^+e^-$ per energy) from the dark matter annihilation. The factor $1/2$ is necessary to not double-count dark matter annihilations.

After taking into account energy losses, the $e^+e^-$ flux recorded at Earth follows then from

\begin{equation}\label{eq: positron flux}
\Phi_e = \frac{c}{4\pi} n_e.
\end{equation}

\subsection{Gamma-Ray Fluxes at Earth}
Furthermore, we keep track of the emitted $\gamma$-ray photons produced by $e^+e^-$ in the energy loss interactions to investigate the impact of stochastic ICS of the $\gamma$-ray flux.

In the stochastic model, the $\gamma$-rays are readily obtained from the energy losses calculated in each interaction, since the ICS energy loss of a $e^+e^-$ corresponds to the energy transferred to the photon. 

On the other hand, the continuous model calculates the energy lost over a time step, which does not correspond to the energy transferred to a specific photon. Therefore, to compute the continuous $\gamma$-ray flux, we use our stochastic model, and at each time step, average over all possible energy losses. This gives the correct $\gamma$-ray spectrum and averages over the $e^+e^-$ energy losses as in the continuous model.

\subsection{Simulation Models}
\label{sec: simulation models}
We compute the $e^+e^-$ spectra for the following 5 dark matter masses: 10~TeV, 30~TeV, 50~TeV, 100~TeV and 300~TeV. Additionally, since the magnetic field strength is not well known, we consider 3 different magnetic field strengths 1~$\mu$G, 2~$\mu$G and 3~$\mu$G, which determine the impact of the synchrotron energy losses. In our standard scenario, we take $m_{\rm DM} = 100$~TeV and ${ B = 1\,\mu }$G. For each data set, we simulate sufficient $e^+e^-$ to achieve statistical accuracy, which corresponds to about $~300\,000$ particles per data set.

In the following, we present local $e^+e^-$ fluxes normalised according to Equations~\ref{eq: positron rate} and~\ref{eq: positron flux}. For illustrative purposes, we choose a dark matter annihilation cross section of ${ \langle\sigma v\rangle = 10^{-24} }$~cm$^3$/s that would produce a signal detectable at the expected CTA effective area and an expected energy resolution of ${\sim 5\% }$~\cite{2013APh....43..171B}. While this cross section is rather large compared to the thermal cross section of ${\sim 3\times10^{-26}}$~cm$^3$/s~\cite{Steigman:2012nb}, we point out that mechanisms such as Sommerfeld enhancement can boost the thermal cross section by several orders of magnitude even at hundreds of TeV, see e.g.~\cite{Lattanzi:2008qa}. However, we note that the flux can easily be re-scaled to a different annihilation cross section by applying Equation~\ref{eq: positron rate}. Our results do not depend on the annihilation cross section, \textit{i.e.}, the relative flux between the stochastic and continuous model does not change.

\section{Results}
\label{sec: results}

\begin{figure*}[tb]
\begin{minipage}[t]{0.48\textwidth}
\centering
\includegraphics[width=0.98\textwidth]{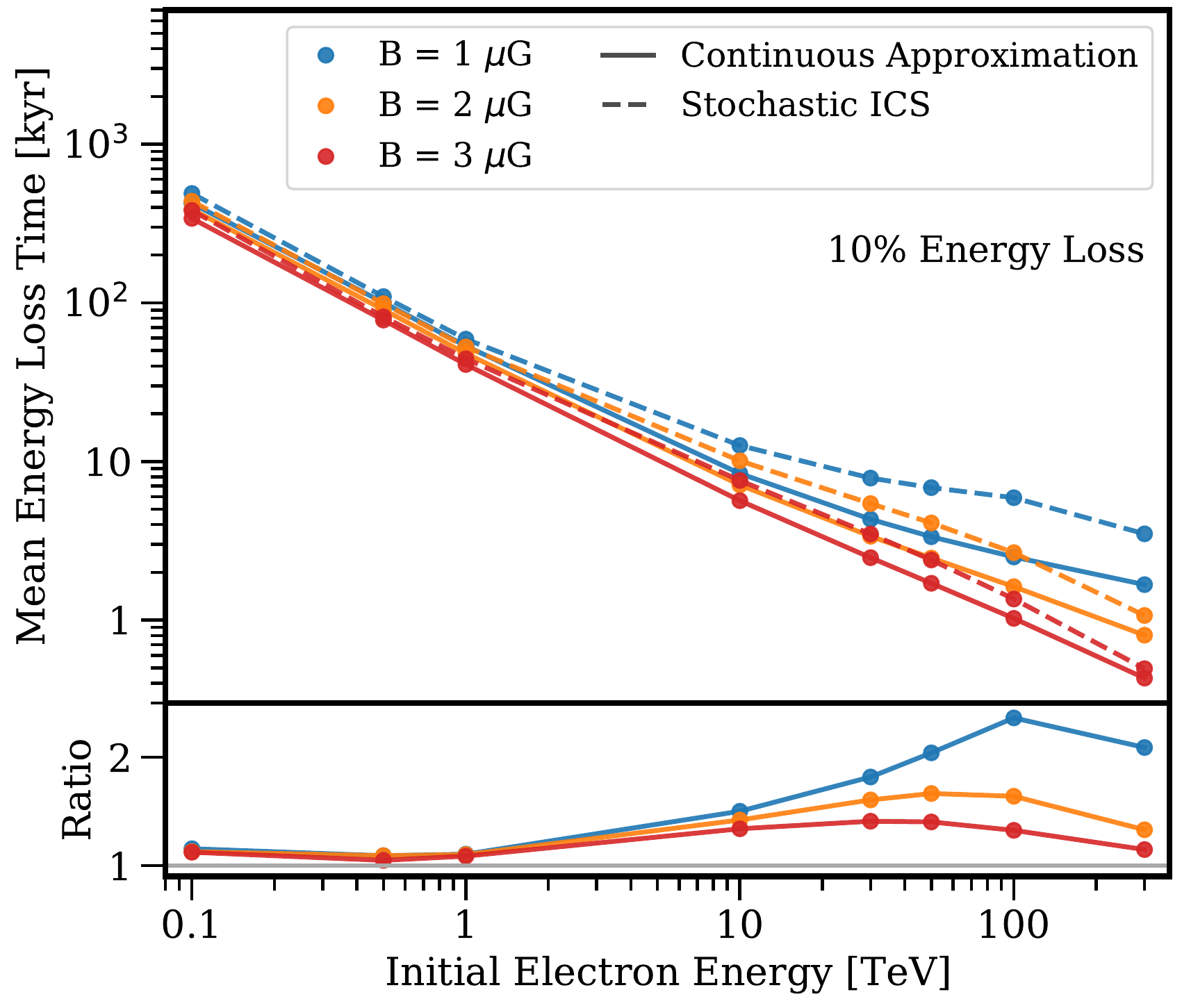}
\end{minipage}
\hfill
\begin{minipage}[t]{0.48\textwidth}
\centering
\includegraphics[width=0.98\textwidth]{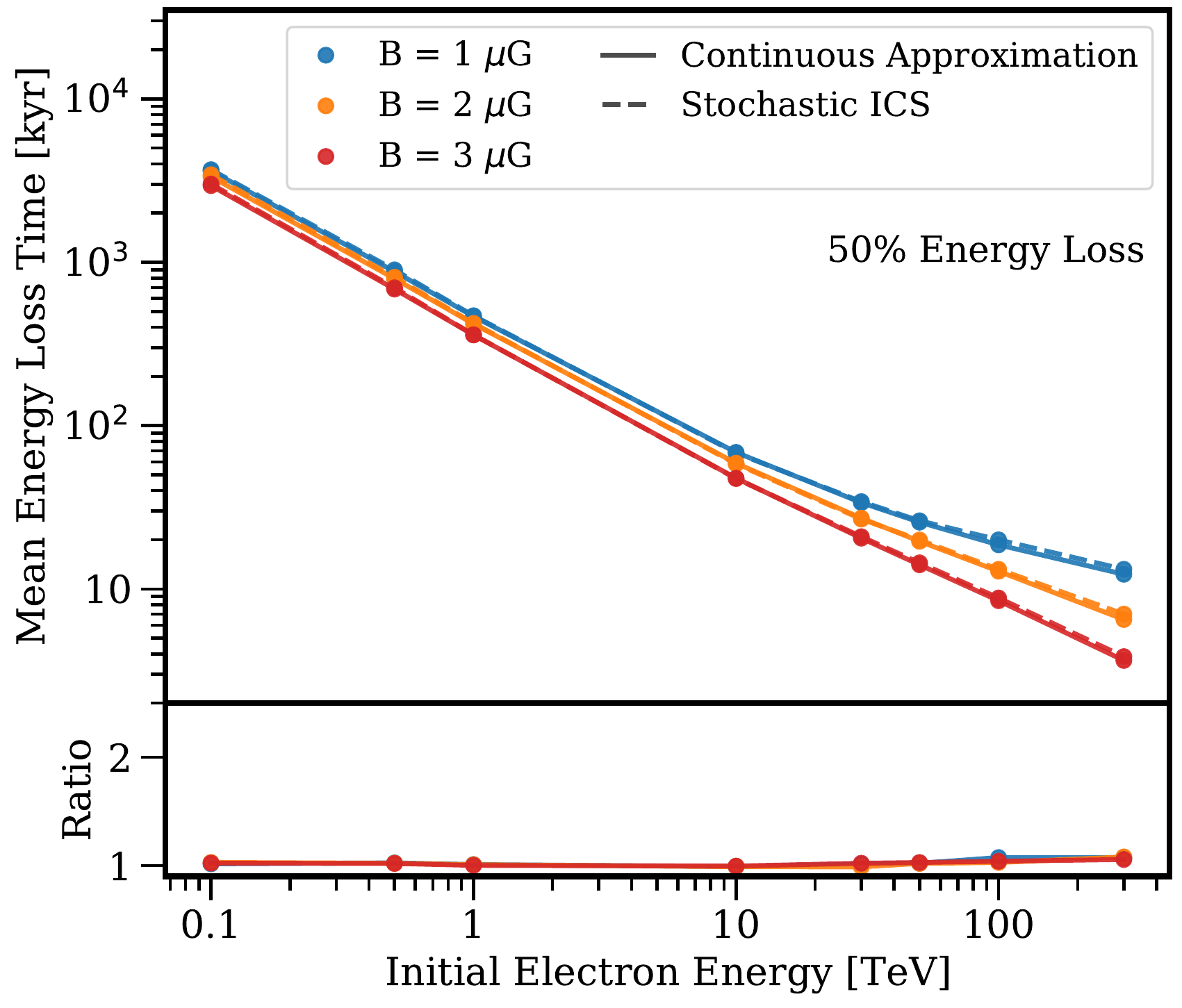}
\end{minipage}
\caption{The mean time it takes an $e^+e^-$ to lose 10\% (left panel) and 50\% (right panel) of its initial energy for the three different magnetic field strengths. Solid lines represent the continuous energy loss model and dashed lines the stochastic ICS model. The bottom panels show the ratio of the stochastic/continuous loss times. In the stochastic case, the low probability of ICS interactions, coupled with the significant energy loss per interaction, means that the mean energy-loss time is longer than in the continuous case. For larger total energy losses (e.g., 50\%), this effect begins to fade and the mean energy loss times approach each other.}
\label{fig: mean loss time}
\end{figure*}

\begin{figure}[tbp]
\centering
\includegraphics[width=0.48\textwidth]{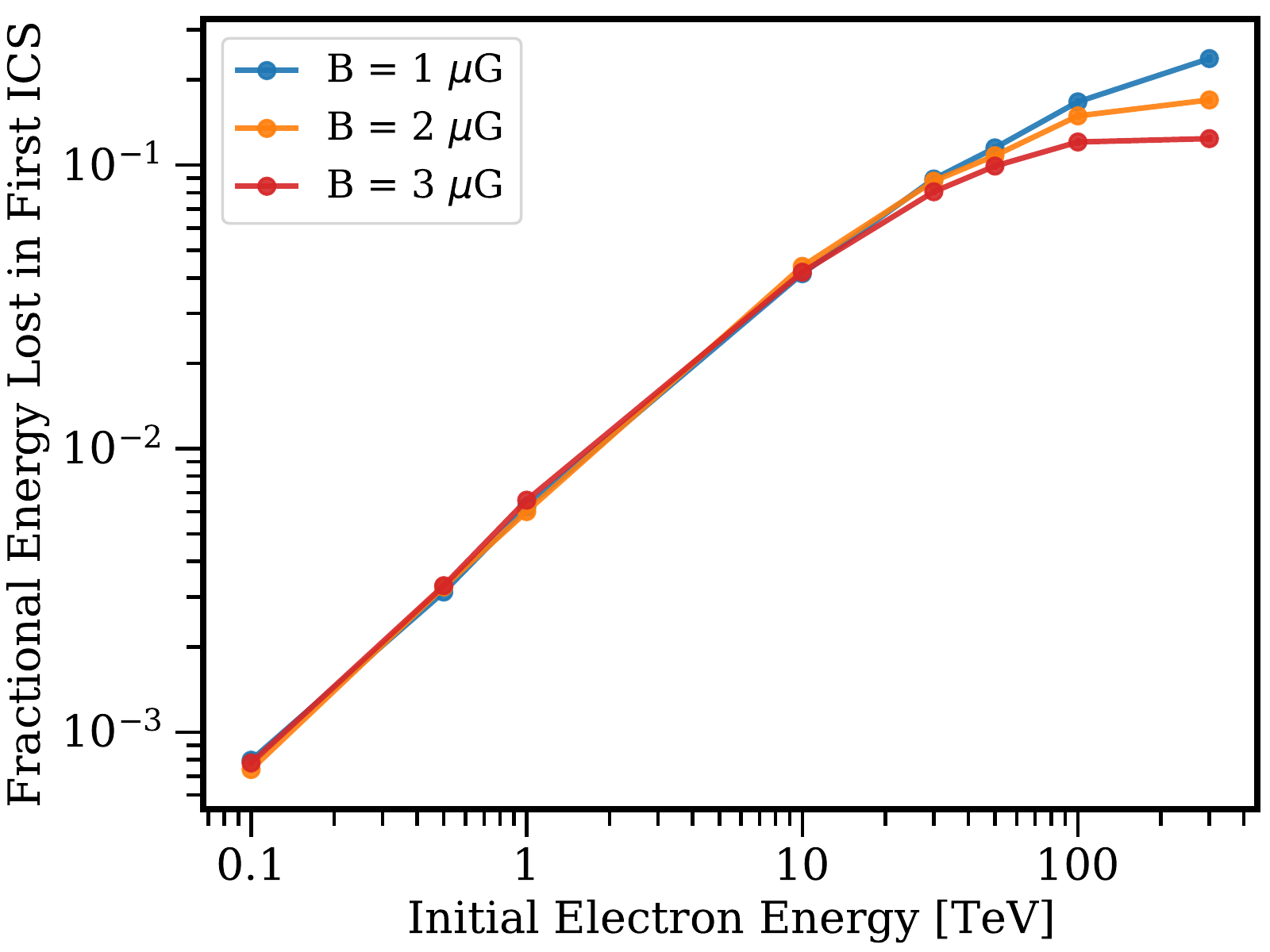}
\caption{The average energy loss in the first ICS interaction after injection as a function of the initial $e^+e^-$ energy, for the three different magnetic field strengths. To observe the enhancement of the dark matter spectral feature experimentally, the energy resolution should roughly be smaller than the energy loss in the first ICS interaction.}
\label{fig: E first ICS vs E init}
\end{figure}

\subsection{Electron and Positron Spectra}
\label{sec: electrons and positrons}
Figure~\ref{fig: fig1} shows a 100~TeV dark matter particle annihilating directly into an $e^+e^-$ pair. The magnetic field strength is ${ B = 1\,\mu }$G. The upper panel shows the combined $e^+e^-$ flux multiplied by $E^2$ as a function of $e^+e^-$ energy, $E$, for the continuous approximation (blue) and the stochastic model (red). The lower panel shows the relative difference between the two models. At energies corresponding to the injection energy of the $e^+e^-$, the flux in the stochastic model exceeds the continuous approximation by a factor of 2.6. For reference, we show an astrophysical background extrapolated from HESS $e^+e^-$  data~\cite{HESSdata} using a simple power law (of course, the true astrophysical flux could be different).

We note that the results shown here (and throughout this paper), are illustrated for a benchmark dark matter annihilation rate of $\sim$10$^{-24}$~cm$^{3}$s$^{-1}$. This parameter is set due to the fact that the enhancement in the dark matter signal due to stochastic effects is independent of the assumed dark matter cross-section. We choose a cross-section of $\sim$10$^{-24}$~cm$^{3}$s$^{-1}$ due primarily to the fact that it produces a dark matter flux that exceeds the extrapolated astrophysical background at energies near the dark matter mass. This means that experimental searches with the required sensitivity should observe the dark matter signal in a low-background environment. As is typical for dark matter searches for strong spectra signatures, the sensitivity of a suitably large acceptance experiment could potentially set constraints far below the cross-sections which provide a S/N ratio exceeding 1 (see for example, Fermi-LAT line searches, which set cross-sections near S/N ratios of 10$^{-3}$~\cite{Foster:2022nva}, but we note that the S/N ratio will increase (perhaps by up to a factor of 10) for the broader post-propagation spectra such as in this work.)

Figures~\ref{fig: 300 TeV 1 muG}~and~\ref{fig: 10 TeV 3 muG} show two more scenarios, representing an optimistic and a pessimistic case, respectively. In Figure~\ref{fig: 300 TeV 1 muG}, the dark matter mass is 300~TeV and the magnetic field strength 1~$\mu$G. The enhancement is 2.2, as stochastic ICS effects are strongest at higher energies, while the effects of synchrotron losses are minimized due to the small magnetic field strengths. Notably, in this case, the energy losses from individual ICS interactions become so large (as even ICS interactions with the CMB approach the Klein-Nishina limit), that we see a substantial dip in the electron spectrum at energies between $\sim$30-200~TeV due to electrons that lose nearly all of their energy in the first ICS interaction.

Conversely, in Figure~\ref{fig: 10 TeV 3 muG}, the dark matter mass is 10~TeV and the magnetic field strength 3~$\mu$G. Here, the enhancement is smaller, but still significant, reaching a peak of 1.4 near the dark matter mass. Figures for other dark matter masses and magnetic field strengths are shown in the Appendix. Across the various cases the enhancement of the local $e^+e^-$ flux due to stochastic ICS at energies within 5\% of the dark matter mass is about a factor of 2, and is higher for larger dark matter masses and weaker magnetic fields strengths.

\begin{figure*}[tb]
\begin{minipage}[t]{0.48\textwidth}
\centering
\includegraphics[width=0.98\textwidth]{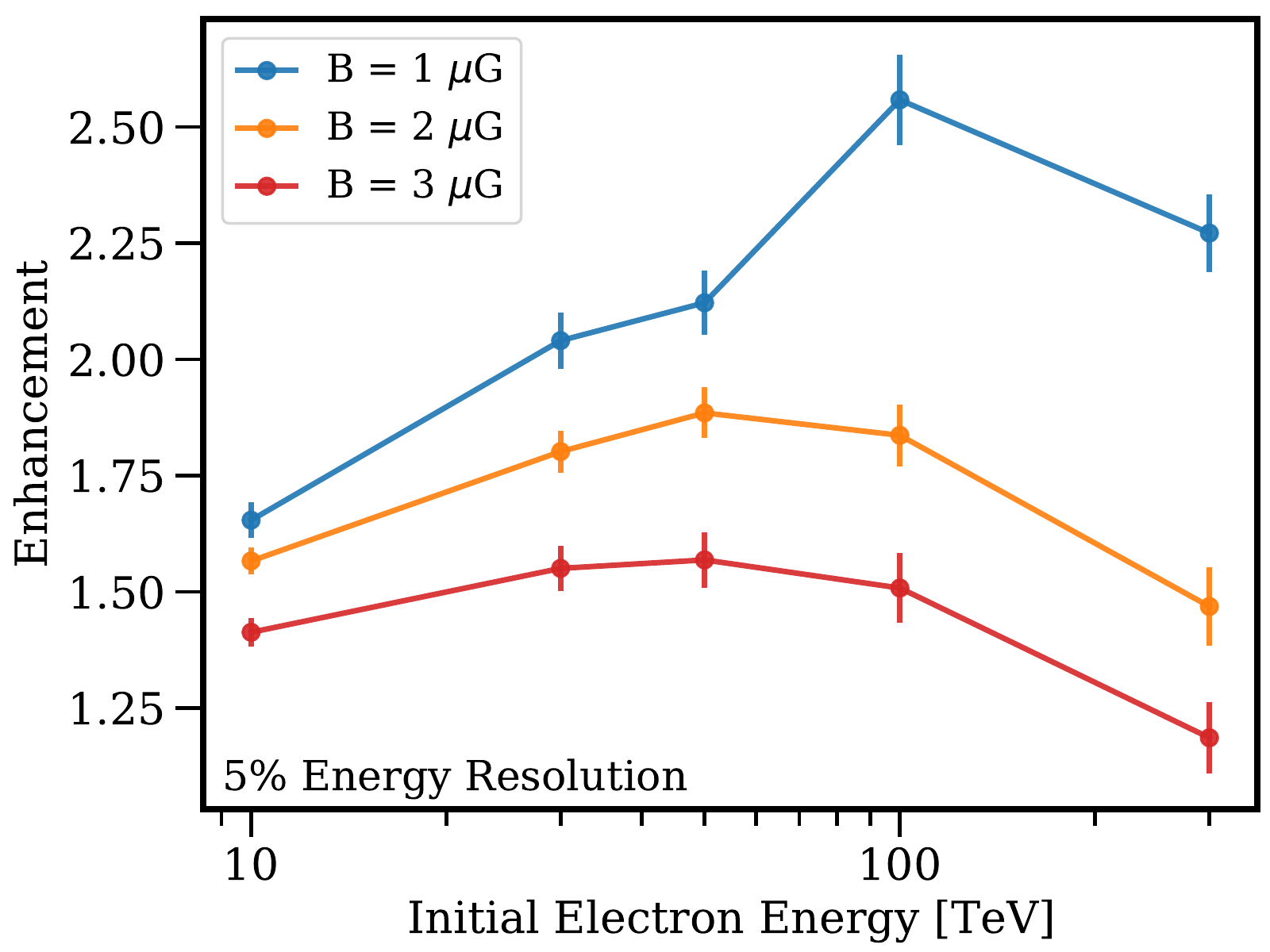}
\caption{The enhancement of the dark matter cutoff in the stochastic loss model compared to the continuous model for the various initial $e^+e^-$ energies, and the three different magnetic field models. The energy resolution is fixed to 5\% -- with a better energy resolution, enhancements at lower initial $e^+e^-$ energies become stronger.}
\label{fig: enhancement vs DMmass}
\end{minipage}
\hfill
\begin{minipage}[t]{0.48\textwidth}
\centering
\includegraphics[width=0.98\textwidth]{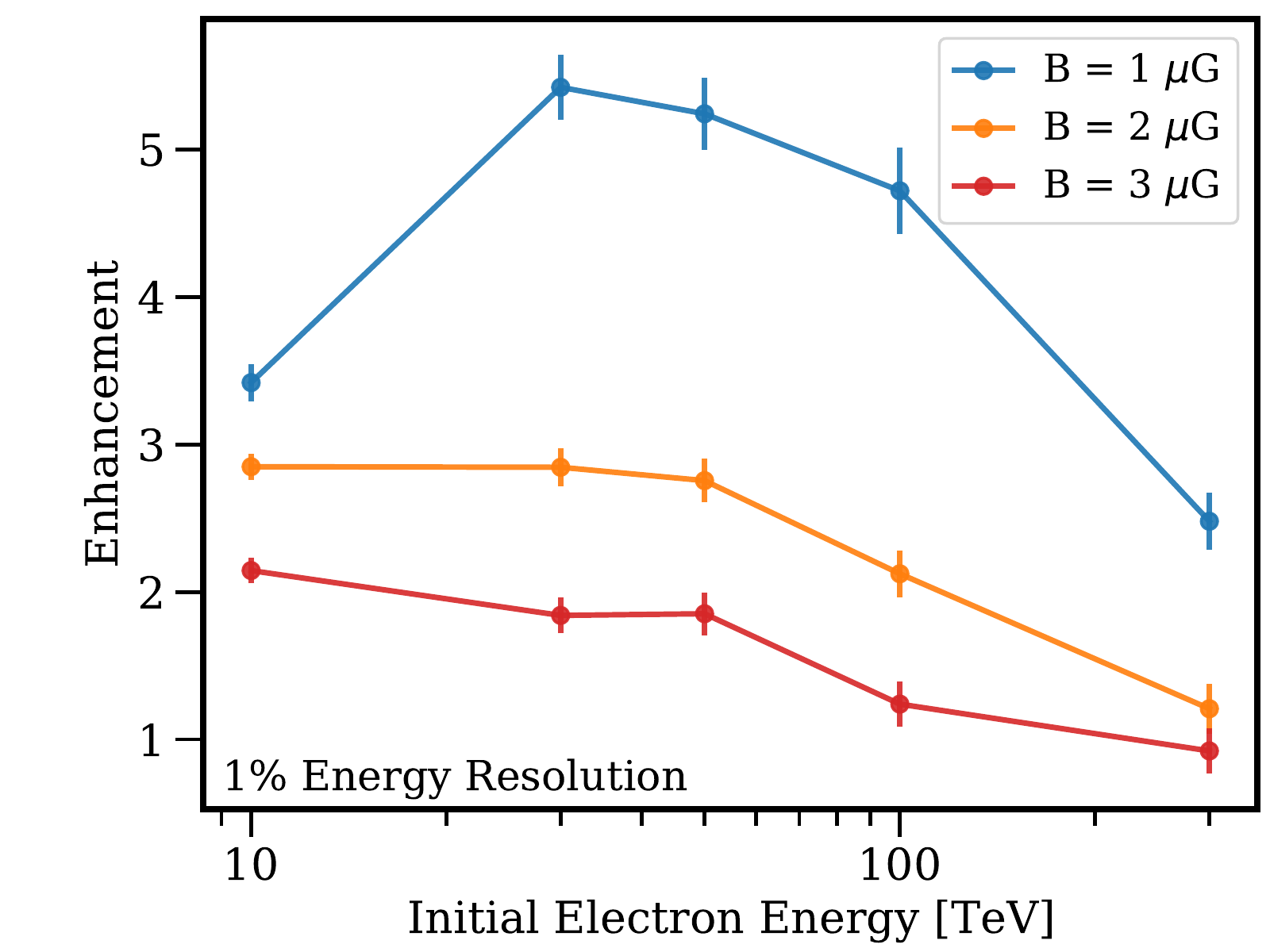}
\caption{The enhancement of the dark matter cutoff in the stochastic loss model compared to the continuous model for the various initial $e^+e^-$ energies, and the three different magnetic field models. The energy resolution is fixed to 1\% -- with a better energy resolution, enhancements at lower initial $e^+e^-$ energies become stronger.}
\label{fig: enhancement vs DMmass 0.01}
\end{minipage}
\end{figure*}

\begin{figure*}[tb]
\begin{minipage}[t]{0.48\textwidth}
\centering
\includegraphics[width=0.98\textwidth]{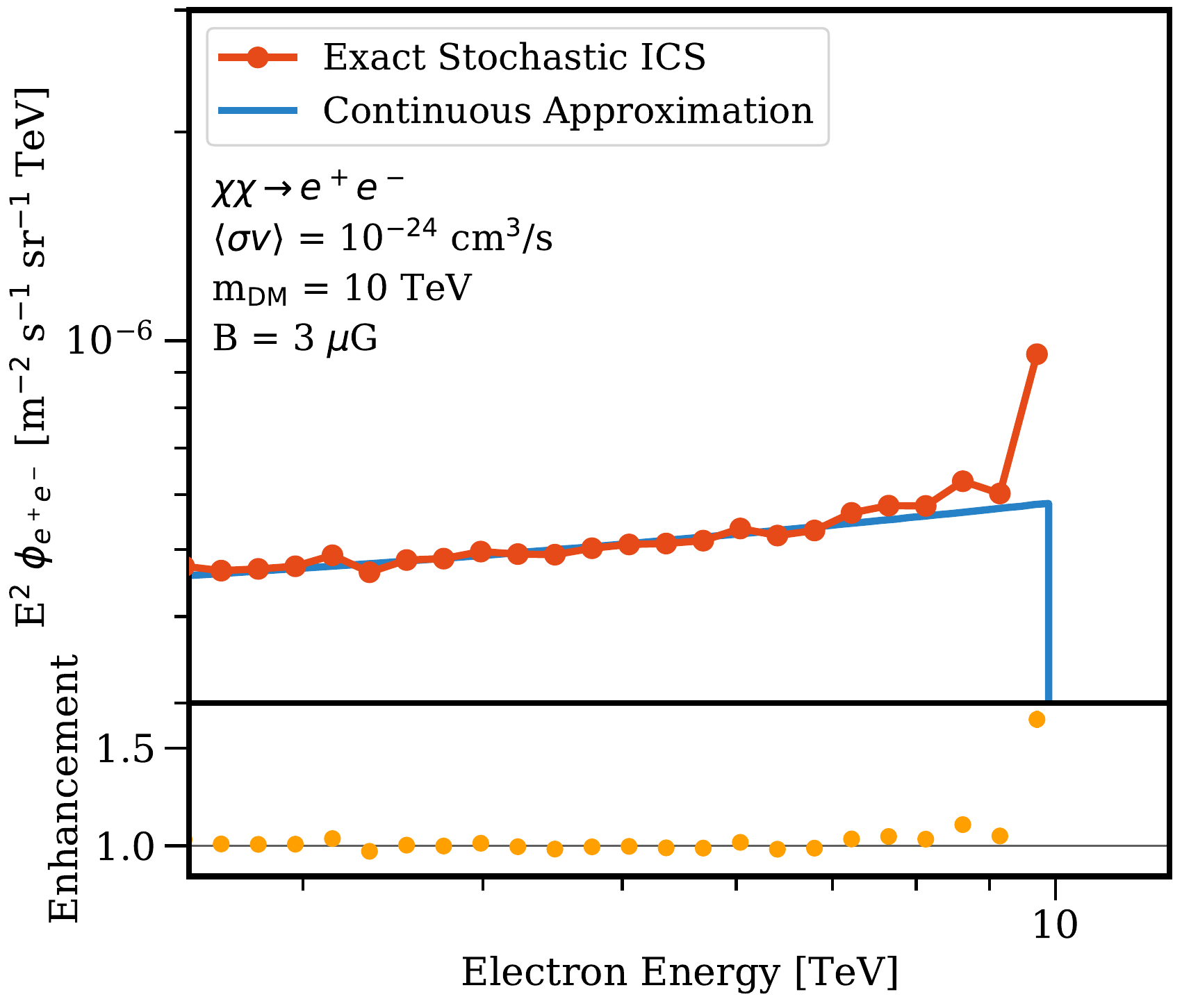}
\caption{The $e^+e^-$ flux expected from an annihilating dark matter particle with a mass of 10~TeV for a magnetic field strength of 3~$\mu$G, similar to Figure~\ref{fig: 10 TeV 3 muG}, but with a smaller energy binning of 3\% instead of 5\%, which improves the resolution of the enhancement in the stochastic model (red) compared to the continuous model (blue), increasing the enhancement from about 1.4 to 1.6.}
\label{fig: 10 TeV 3 muG 0.03}
\end{minipage}
\hfill
\begin{minipage}[t]{0.48\textwidth}
\centering
\includegraphics[width=0.98\textwidth]{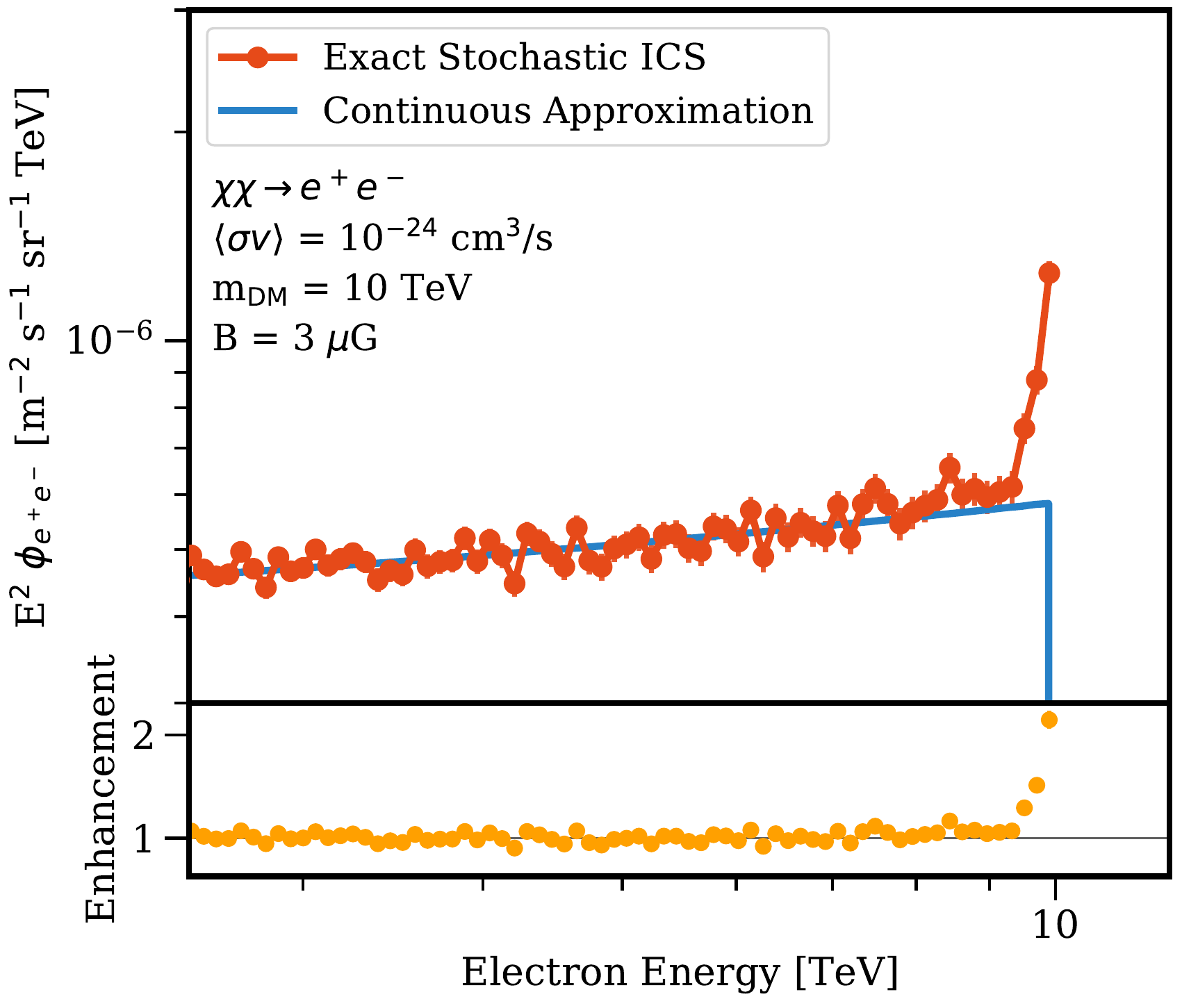}
\caption{The $e^+e^-$ flux expected from an annihilating dark matter particle with a mass of 10~TeV for a magnetic field strength of 3~$\mu$G, similar to Figure~\ref{fig: 10 TeV 3 muG}, but with a smaller energy binning of 1\% instead of 5\%, which improves the resolution of the enhancement in the stochastic model (red) compared to the continuous model (blue), increasing the enhancement from about 1.4 to 2.1.}
\label{fig: 10TeV 3 muG 0.01}
\end{minipage}
\end{figure*}

Figure~\ref{fig: mean loss time} shows the mean time it takes an $e^+e^-$ to lose some fraction of its initial energy as a function of its initial energy (\textit{i.e.}, the dark matter mass). The left panel shows the mean energy loss time for a 10\% energy loss, and the right panel for a 50\% energy loss. It can be seen that the mean energy loss times are significantly longer (about a factor of 2 for a 10\% energy loss) in the stochastic model compared to the continuous model, and this ratio is larger for larger injection energies. Additionally, the mean energy loss time in both the continuous and stochastic cases increases with decreasing magnetic field strength, as this reduces energy losses from synchrotron interactions. We note that, within the stochastic case, it is possible that the 10\% energy loss occurs entirely due to synchrotron radiation before any ICS interactions have occurred. For the 50\% energy loss case, the stochastic and continuous loss times are closer compared to the 10\% energy loss case and only differ by about a factor of 3\% at 100~TeV. This is due to the fact that the effect of stochasticity is smoothed out once multiple ICS interactions are likely.

Figure~\ref{fig: E first ICS vs E init} shows the fractional energy lost in the first ICS interaction (\textit{i.e.}, the energy lost in the first ICS interaction divided by the initial energy) for the various initial $e^+e^-$ energies and magnetic field strengths. This is an indication of the energy resolution required for telescopes to observe the enhanced feature. For smaller dark matter masses, the fractional energy lost becomes smaller, and a better energy resolution is required. For an $e^+e^-$ with an initial energy of 0.1~TeV, the energy loss in the first ICS interaction is about 0.1~GeV, which would require an energy resolution of at least 0.1\% to be experimentally observable. On the other hand, for an initial energy of 100~TeV, the energy loss is about 10~TeV, which corresponds to an energy resolution of 10\%, which is feasible for upcoming experiments.

We note that Figures~\ref{fig: mean loss time} and~\ref{fig: E first ICS vs E init} justify our choice to ignore the effect of cosmic-ray diffusion on the local $e^+e^-$ spectrum throughout this work. In particular, the time until an $e^+e^-$ with an initial energy of 100~TeV loses 50\% of its energy is $t \sim$~70~kyr. The average spatial displacement at this energy is given by $L = \sqrt{6Dt}$, where $D$ is the diffusion coefficient given by $D = D_0 \left(E\right)^\delta$. Using a typical normalization factor of $D_0 = 2\times10^{28}$~cm$^2$/s at 1~GeV, and a diffusion index $\delta = 0.4$~\cite{Hooper:2017gtd, Korsmeier:2021bkw}, a 100-TeV $e^+e^-$ displaces about a distance of $\sim$~1700~pc. This is well below the Galactic halo size of $\sim$5~kpc, within which cosmic rays are contained, and also much smaller than the radius over which the dark matter density significantly varies~\cite{Navarro:1995iw}. Even for a 100~GeV $e^+e^-$, where the time required to lose 50\% of the initial energy is $\sim 3600$~kyr, the average displacement becomes $\sim$~3~kpc. Even if some $e^+e^-$ escape the Galaxy, this happens at energies well below the injection energy, where our effect is observed. 

\begin{figure*}[tb]
\begin{minipage}[t]{0.48\textwidth}
\centering
\includegraphics[width=0.98\textwidth]{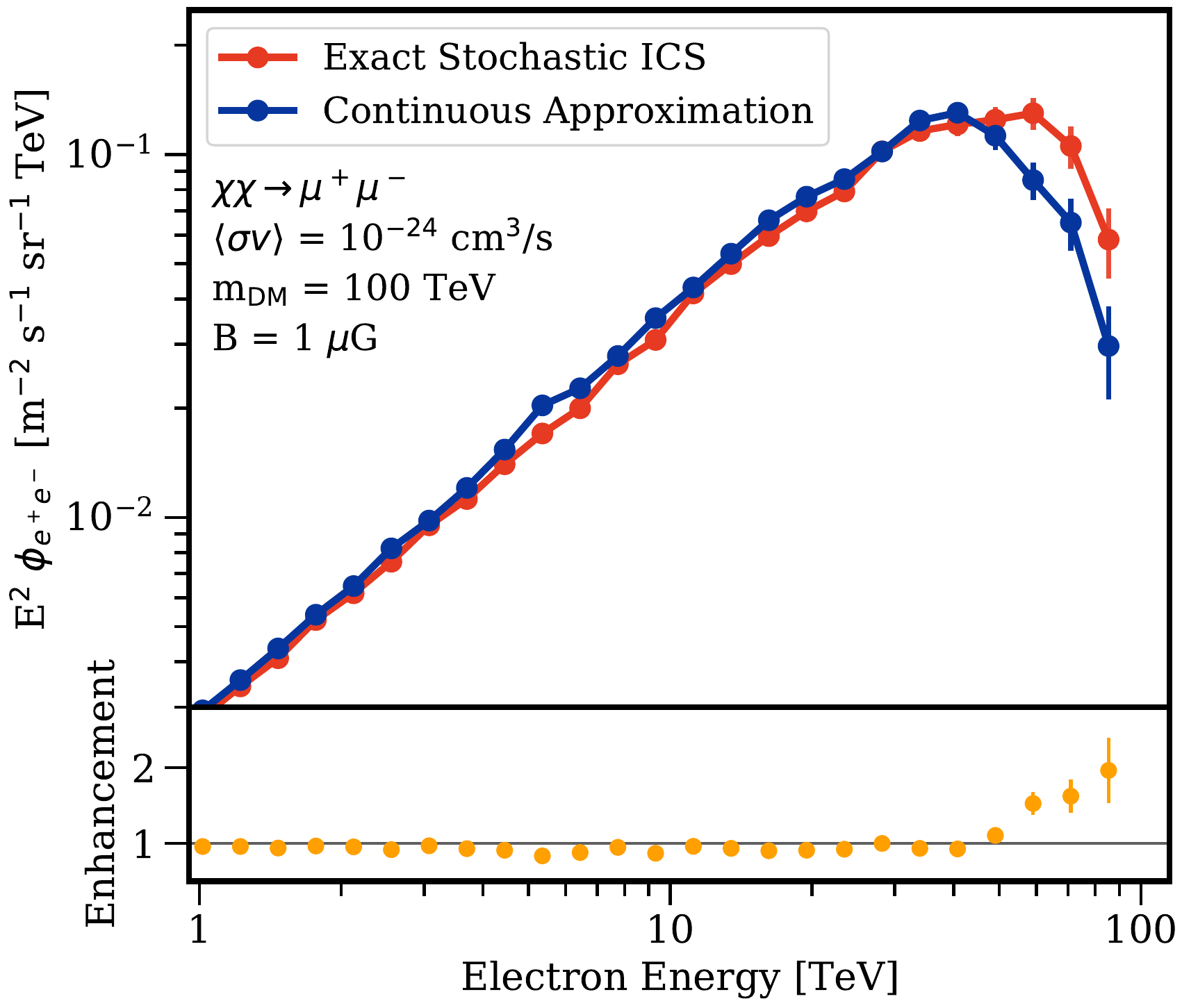}
\caption{The $e^+e^-$ spectrum from a 100~TeV dark matter particle that annihilates into $\mu^+\mu^-$, assuming a magnetic field strength of 1~$\mu$G. For energies above $\sim$50~TeV, the stochastic model (red) is enhanced by about a factor of 2 compared to the continuous approximation (blue).}
\label{fig: muon 100 TeV}
\end{minipage}
\hfill
\begin{minipage}[t]{0.48\textwidth}
\centering
\includegraphics[width=0.98\textwidth]{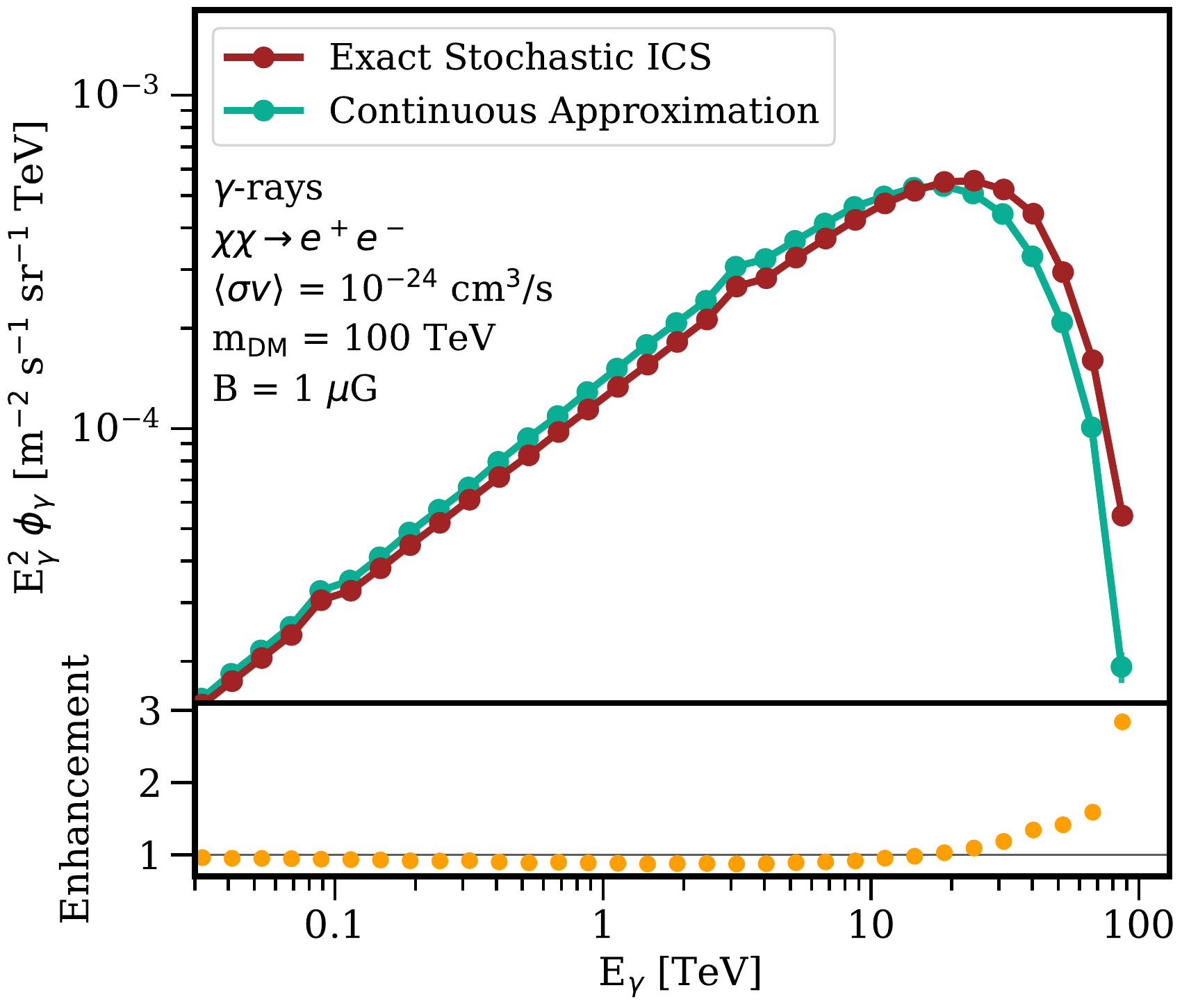}
\caption{The $\gamma$-ray flux as a function of $\gamma$-ray energy from the annihilation of a 100~TeV dark matter particle into $e^+e^-$ pairs that emit $\gamma$-rays when cooling, assuming a magnetic field strength of 1~$\mu$G. At the highest energies, the stochastic energy loss model (red) shows an enhancement of about a factor of 3 compared to the continuous energy loss model (blue).}
\label{fig: Gamma 100 TeV 2 muG}
\end{minipage}
\end{figure*}

In Figure~\ref{fig: enhancement vs DMmass} we summarize our results for the various dark matter masses and show the enhancement of the spectral cutoff in the stochastic model compared to the continuous model for the different magnetic field strengths. We choose an energy resolution of 5\%, since this is a realistic value for upcoming experiments, but note that the enhancement for lower initial energies would be stronger for even better energy resolution. In Figure~\ref{fig: enhancement vs DMmass 0.01}, we show such a scenario, where the energy resolution is instead set to 1\%. This provides an enhancement of more than a factor of 2 even at 10~TeV across all magnetic field models (see also Figure~\ref{fig: E first ICS vs E init}). Additionally to further demonstrate the dependence of the enhancement on the energy resolution, Figures~\ref{fig: 10 TeV 3 muG 0.03} and~\ref{fig: 10TeV 3 muG 0.01} show an alternative to Figure~\ref{fig: 10 TeV 3 muG}, where we refine the energy binning from 5\% to 3\% and 1\%, respectively. This increases the enhancement from a factor of 1.4 to 1.6 and 2.1, respectively, showing that a significant enhancement to the e$^+$e$^-$ flux can also be observed below 100~TeV, given an experiment with sufficient energy resolution.

\subsection{Local e$^+$e$^-$ Spectrum from Dark Matter Annihilation to Muons}
We also consider the case where dark matter particles annihilate into a $\mu^+\mu^-$ pair that subsequently decays into $e^+e^-$. This smears out the initial $e^+e^-$ injection spectrum from a delta function to a function that is approximately constant in $\frac{dN}{dE}$. Figure~\ref{fig: muon 100 TeV} shows the $e^+e^-$ spectrum for a 100~TeV dark matter particle and a magnetic field strength of 1~$\mu$G. Since the $e^+e^-$ are now injected at a distribution of initial energies, rather than the same initial energy, the enhancement of the feature is smaller, about a factor of 2, in the stochastic model compared to the continuous model. Additionally, the peak of the flux is shifted to higher energies in the stochastic model from about 40~TeV to 60~TeV, which is important when determining the dark matter mass given an observed signal.

For the case of the dark matter annihilating into a $\tau^+\tau^-$ pair (or a hadronic final state) that subsequently produces $e^+e^-$, the difference between the stochastic and continuous model would be even more reduced than in the $\mu^+\mu^-$ case, since the $e^+e^-$ injection spectrum would be even softer.

\subsection{Gamma-Ray Spectra}
\label{sec: gammas}
When $e^+e^-$ undergo an ICS interaction, the ISRF photon is converted to a high-energy $\gamma$-ray with an energy identical to the energy lost by the $e^+e^-$. Figure~\ref{fig: Gamma 100 TeV 2 muG} shows the $\gamma$-ray flux expected from $e^+e^-$ produced by a 100~TeV dark matter particle for the stochastic model (red) and the continuous model (blue) as a function of $\gamma$-ray energy. Similar to the $e^+e^-$ spectra, the stochastic model shows an enhancement at the highest energies of almost a factor of 3. This is expected since in the stochastic model, a larger fraction of the $e^+e^-$ energy can be lost in a single interaction compared to the continuous model, which means that a larger amount of energy is transferred to the photon when the $e^+e^-$ has an energy near its initial value, resulting in higher energetic $\gamma$-rays. We note that while the relative enhancement of the $\gamma$-ray flux is similar to the enhancement in the local $e^+e^-$ spectrum, the largest enhancement does not occur near the energetic peak of the $\gamma$-ray emission, making this effect harder to observe practically with $\gamma$-ray telescopes. However, this shift in the peak of the flux from about 16~TeV in the continuous model to about 24~TeV in the stochastic model can be interpreted as a shift in the dark matter mass, which is relevant when determining the dark matter mass from an experimental signal.

\section{Discussion and Conclusions}
\label{sec: discussion}
In this paper, we have shown the importance of precisely treating energy losses from inverse-Compton interactions as a stochastic process instead of approximating them as a continuous process. Specifically, we have shown the impact of stochastic ICS losses on the expected signal from the annihilation of TeV-scale dark matter particles into $e^+e^-$ pairs. When taking the stochasticity of ICS into account, the sharp cutoff at the dark matter mass is increased by about a factor of 2 compared to the continuous model. This effect is significant for TeV dark matter models that annihilate into $e^+e^-$ pairs across a variety of magnetic field models.

This implies that the detectability of dark matter signals in local $e^+e^-$ measurements is significantly greater than previously expected. This is important for current and near-future experiments, such as CTA~\cite{CTAwebsite, Knodlseder:2020onx, 2013APh....43..171B, maier2019performance}, that are able to observe the local $e^+e^-$ fluxes up to $\sim$100~TeV with energy resolutions of $\sim$10\%. Our results are also relevant, but subdominant, at energies corresponding to the 1.4~TeV $e^+e^-$ excess detected by DAMPE~\cite{DAMPE:2017fbg}. We find that the amplitude of the $e^+e^-$ feature would be enhanced by approximately 6\%. This relatively small enhancement is due to the rather wide (10\%) binning of the $e^+e^-$ data by DAMPE. Several studies have indicated that a nearby dark matter clump would be capable of producing the DAMPE signal~\cite{Berezinsky:2014wya, Hutten:2016jko, Yang:2017cjm, Coogan:2019uij}, a constraint that likely remains robust in light of the relatively small enhancement that our model produces at energies near 1~TeV given the DAMPE energy resolution.

We note that, particularly in the case of atmospheric Cherenkov telescopes, such as HESS and the CTA, that the sharp spectral feature in the e$^+$e$^-$ flux must be detected against a background that not only includes e$^+$e$^-$ from non-annihilation sources, but also from backgrounds in cosmic-ray hadrons and $\gamma$-rays that have been misclassified as e$^+$e$^-$. While hadronic backgrounds can be relatively efficiently removed based on the differences between leptonic and hadronic shower propagation through the atmosphere, the dominance of hadronic over leptonic TeV cosmic-rays means that hadronic leakage can be a significant challenge~\cite{2004NIMPA.516..511B}. On the other hand, $\gamma$-rays also produce leptonic showers, and thus the most effective method of minimizing the $\gamma$-ray background is to point at regions in the sky where the $\gamma$-ray flux is minimal~\cite{Parsons:2016bdk}, though more detailed methods might be possible~\cite{2017PhDT........75E}. The details of this background subtraction sensitively depend on the calibration of each instrument, and lie beyond the scope of this paper. However, we do note that because there is not expected to be a strong corresponding spectral signal in either hadronic cosmic-rays or $\gamma$-rays (save a weak $\gamma$-ray contribution from dark matter induced final state radiation), these backgrounds are primarily statistical in nature, representing an additional smooth background component, rather than representing an irreducible systematic error~\cite{Linden:2013mqa}. Finally, we note that such errors will be minimized in space-based instruments, which offer superior particle discrimination.

Our results are relevant for any dark matter models that directly annihilate at least partially into $e^+e^-$ pairs. Enhancements from annihilations into $\mu^+\mu^-$ exist but are less significant. Additionally, slightly more energetic $\gamma$-rays are expected in our stochastic models. However, this enhancement happens at $\gamma$-ray energies that are above the peak of the $\gamma$-ray emission, and may be difficult to detect.

This result is particularly interesting in light of our recent work in Ref.~\cite{John:2022asa}, which found that the stochasticity of ICS smoothed out peaks in the local $e^+e^-$ spectrum from pulsars. This difference results from the fact that the $e^+e^-$ injection from pulsars is highly peaked in time, but spectrally smooth, while the $e^+e^-$ injection from dark matter is smooth in time, but spectrally peaked. Intriguingly, this breaks the degeneracy between the spectrally peaked features expected from dark matter annihilation and pulsar activity, re-affirming the status of a sharp feature in the local $e^+e^-$ spectrum as unambiguous evidence for dark matter annihilation.

\section*{Acknowledgements}
I.J. and T.L. would like to thank Ilias Cholis and Dan Hooper for an enlightening conversation that motivated the work in this manuscript. I.J. and T.L. are supported in part by the Swedish Research Council under contract 2019-05135, T.L. is also supported by the Swedish Research Council under contract 2022-04283, the Swedish National Space Agency under contract 117/19 and the European Research Council under grant 742104. This project used computing resources from the Swedish National Infrastructure for Computing (SNIC) under project Nos. 2021/3-42, 2021/6-326, 2021-1-24 and 2022/3-27, partially funded by the Swedish Research Council through grant no. 2018-05973.

\bibliography{main}

\appendix

\section{Additional Plots}
\label{app: additional plots}
\begin{figure*}
\centering
\begin{minipage}[t]{0.48\textwidth}
\includegraphics[width=0.98\textwidth]{Energy_spectrum_heavyDM_10TeV_3muG.pdf} 
\end{minipage}
\hfill
\begin{minipage}[t]{0.48\textwidth}
\includegraphics[width=0.98\textwidth]{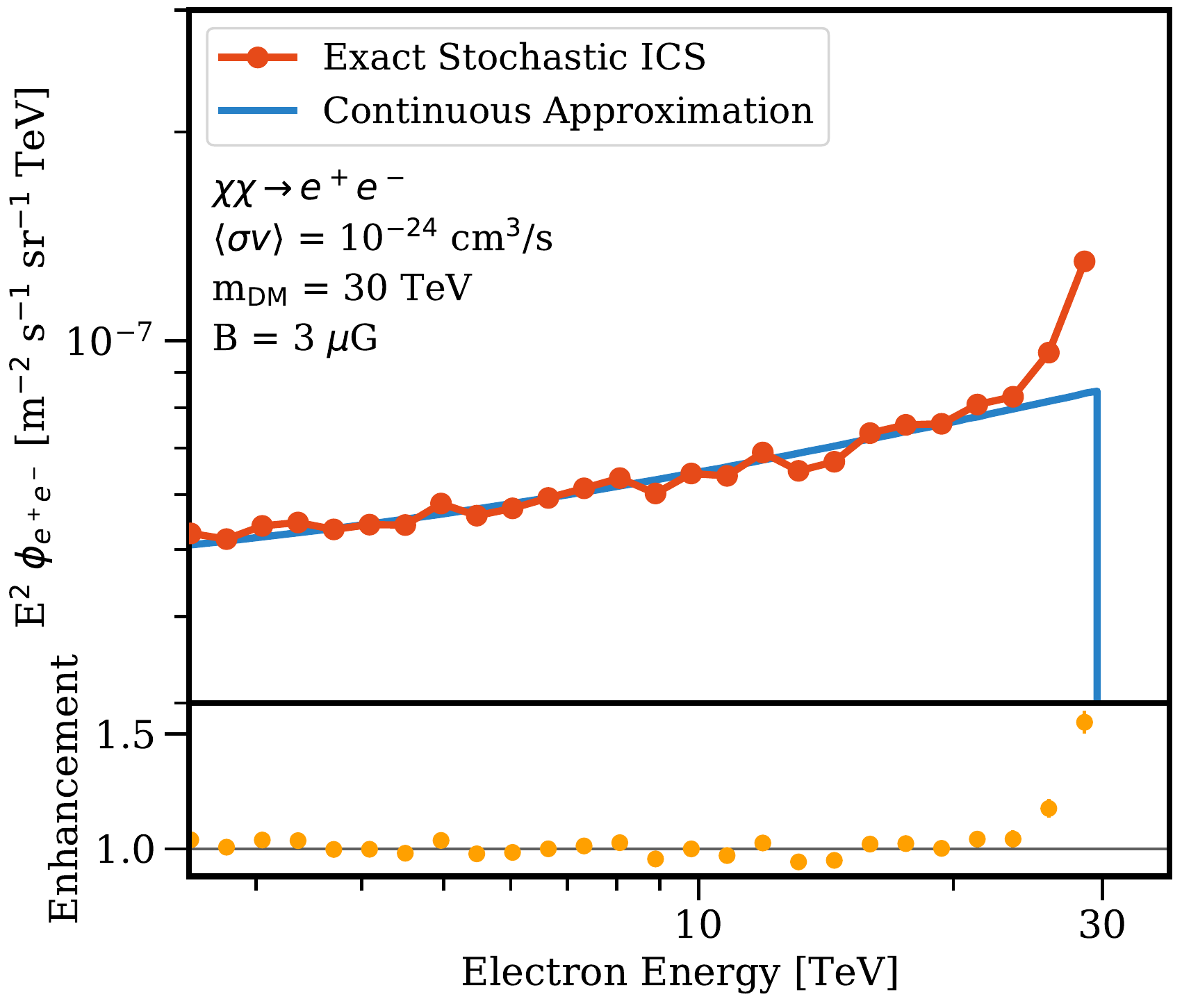} 
\end{minipage}
\vfill
\begin{minipage}[t]{0.48\textwidth}
\includegraphics[width=0.98\textwidth]{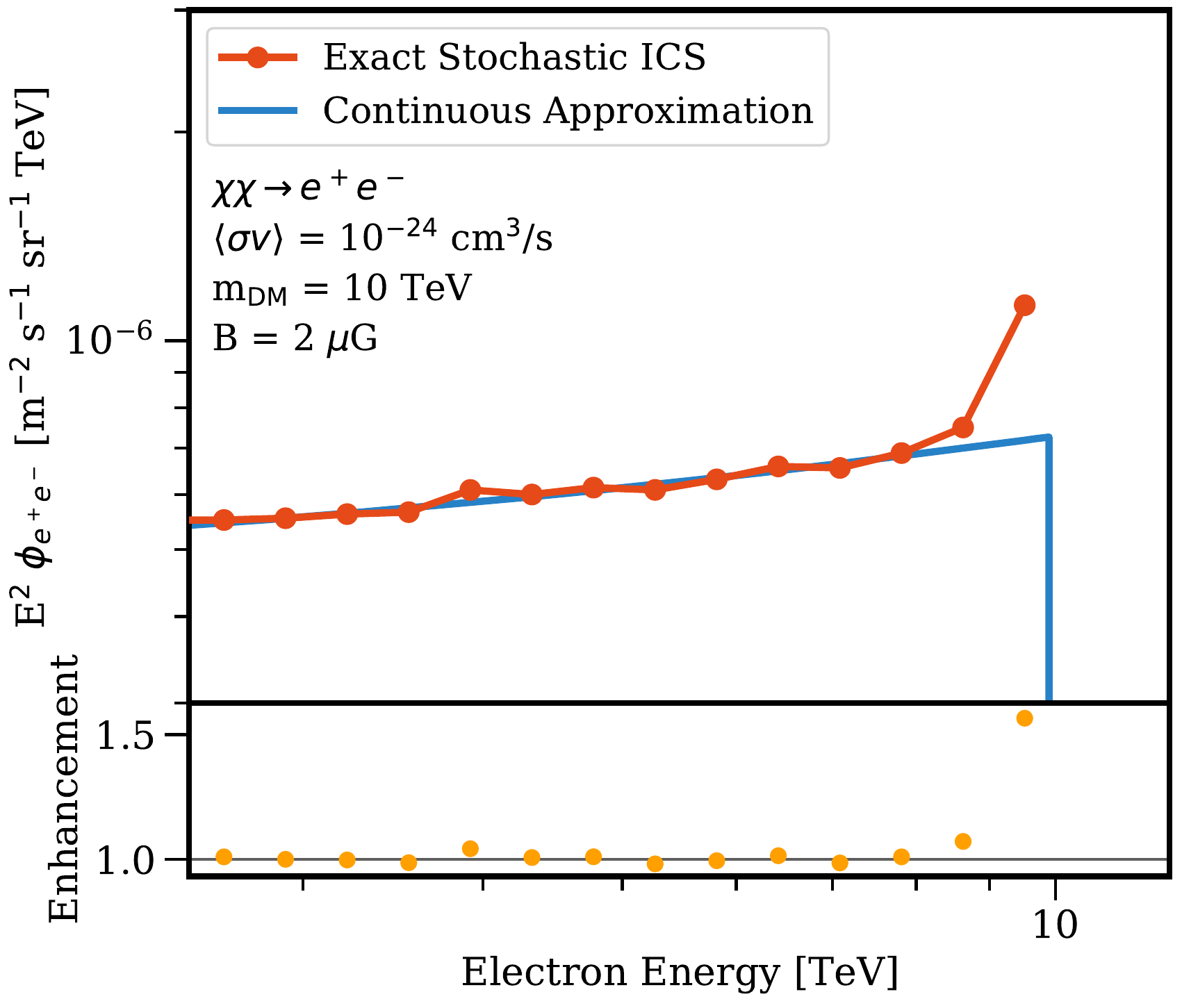} 
\end{minipage}
\hfill
\begin{minipage}[t]{0.48\textwidth}
\includegraphics[width=0.98\textwidth]{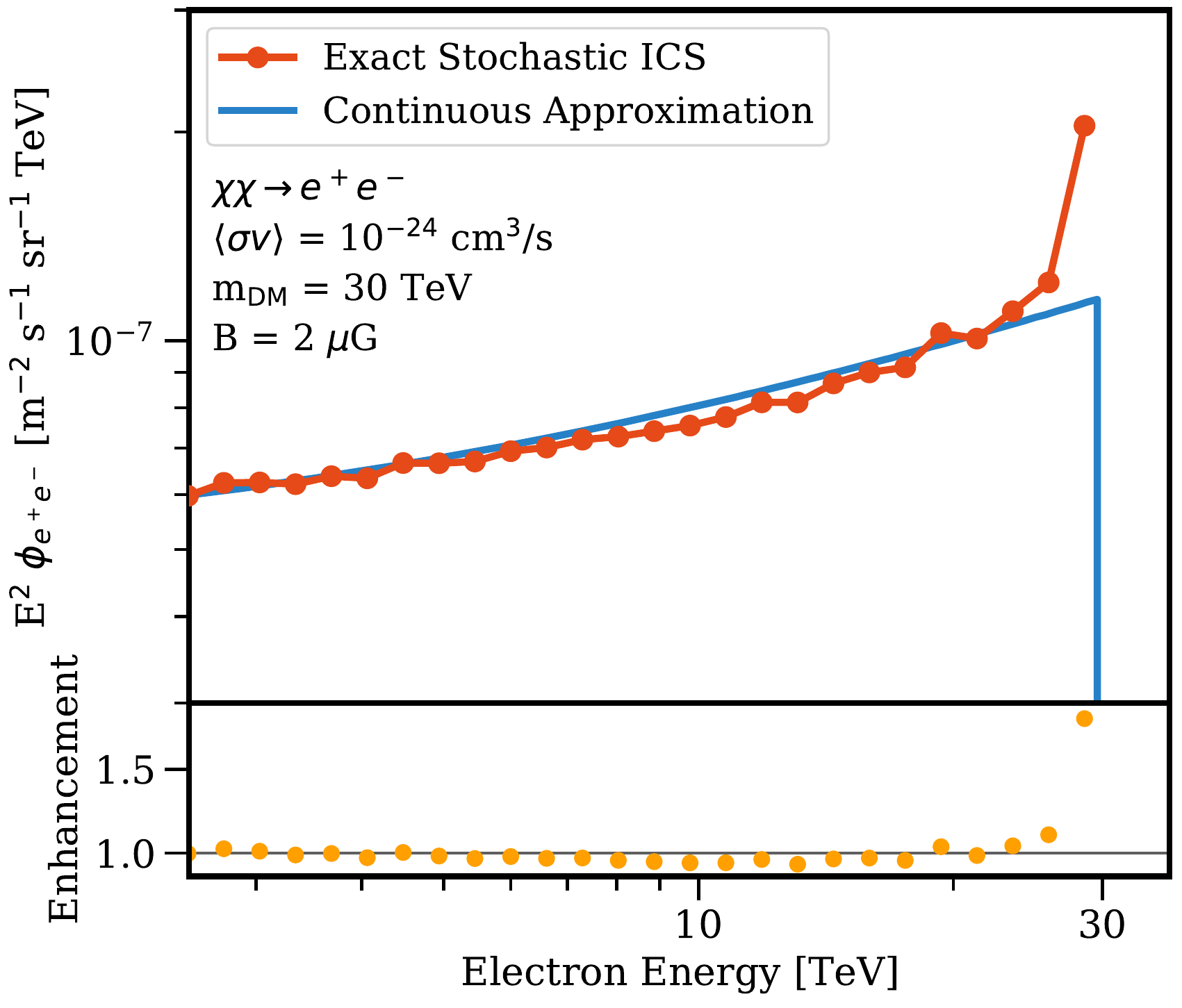} 
\end{minipage}
\vfill
\begin{minipage}[t]{0.48\textwidth}
\includegraphics[width=0.98\textwidth]{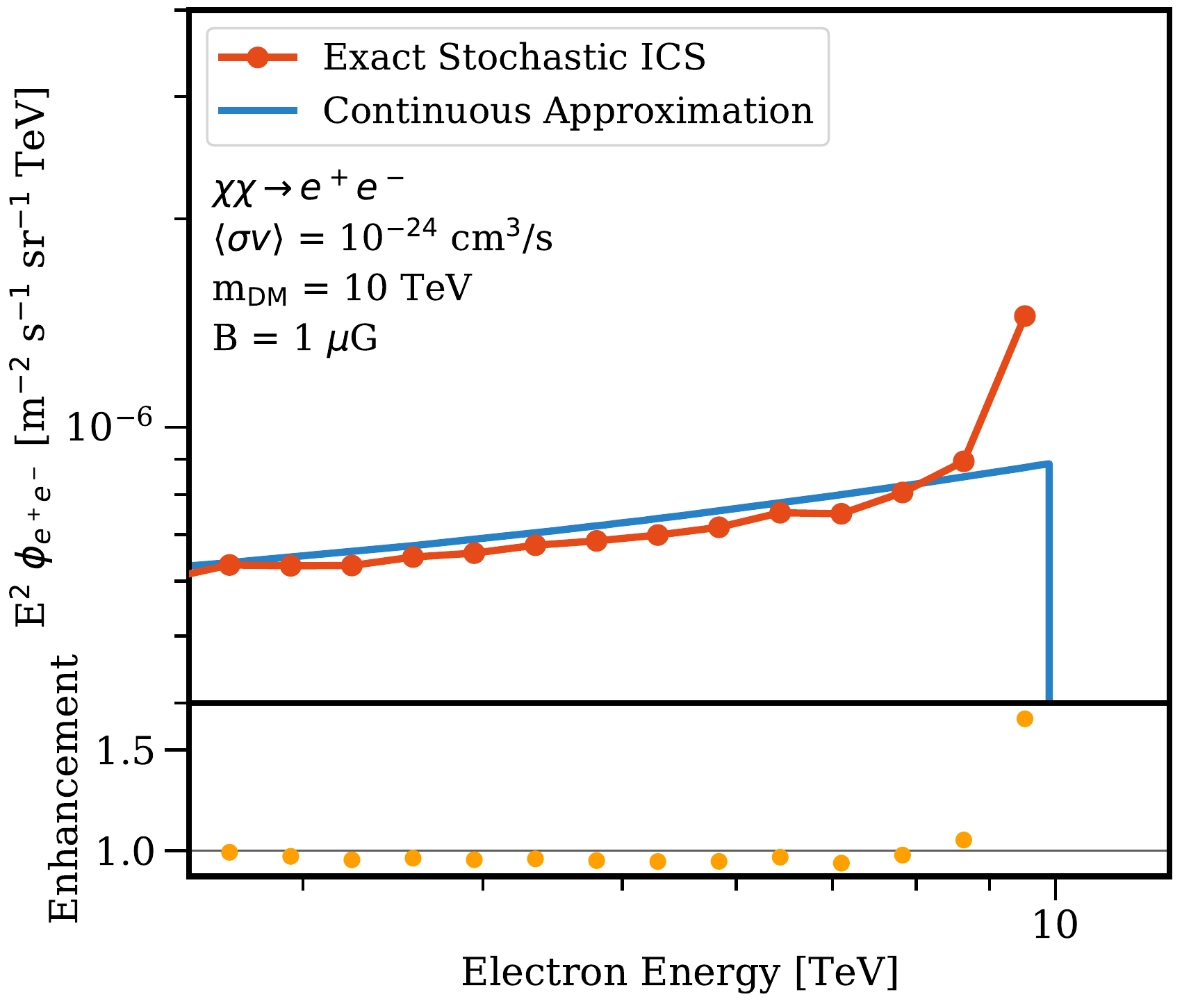} 
\end{minipage}
\hfill
\begin{minipage}[t]{0.48\textwidth}
\includegraphics[width=0.98\textwidth]{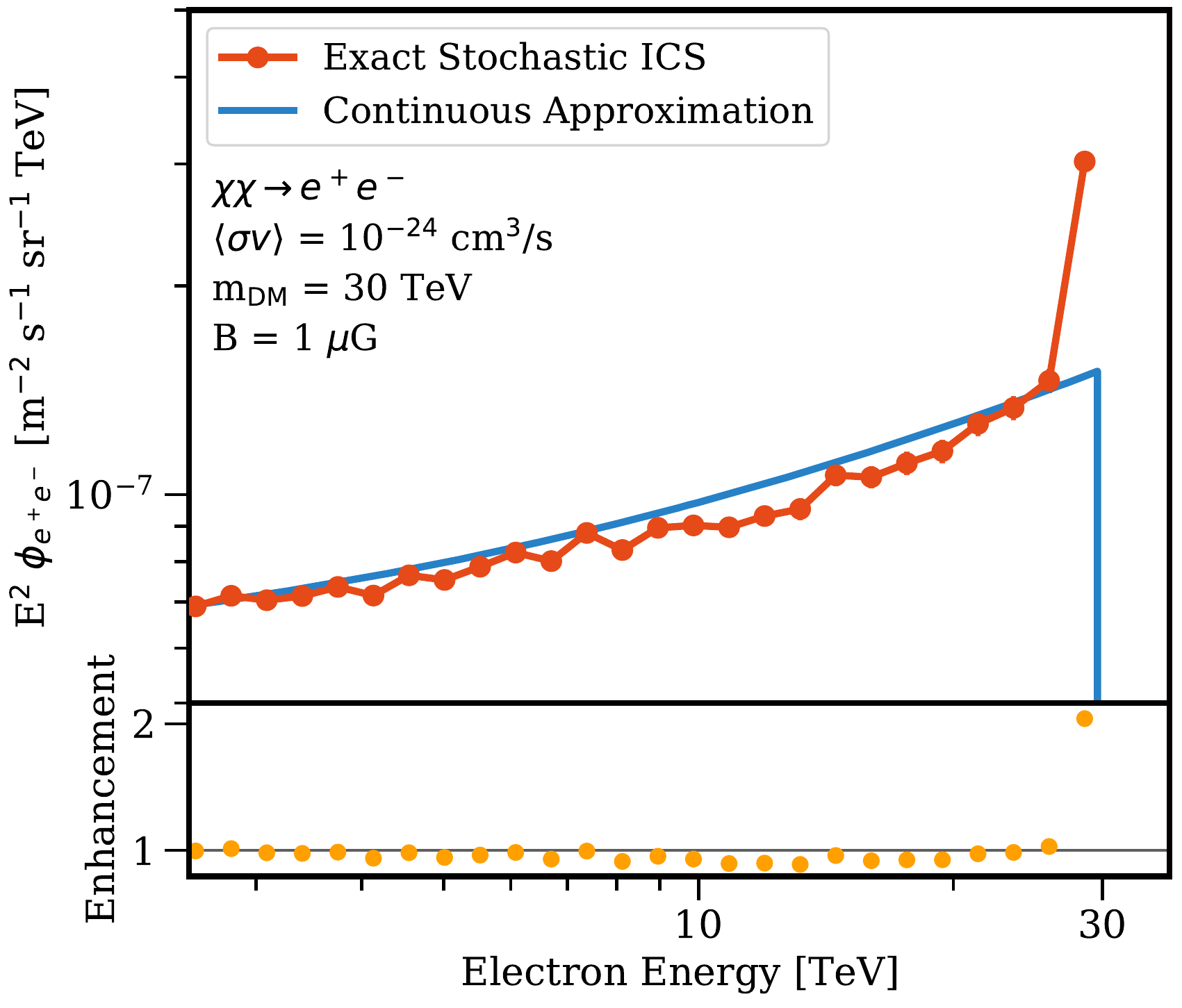} 
\end{minipage}
\caption{The expected $e^+e^-$ flux from the stochastic model (red) compared to the continuous approximation (blue), for a dark matter mass of 10~TeV (left panels) and 30~TeV (right panels) for the three different magnetic field strengths.}
\label{fig: 10 TeV 30 TeV}
\end{figure*}

\begin{figure*}
\centering
\begin{minipage}[t]{0.48\textwidth}
\includegraphics[width=0.98\textwidth]{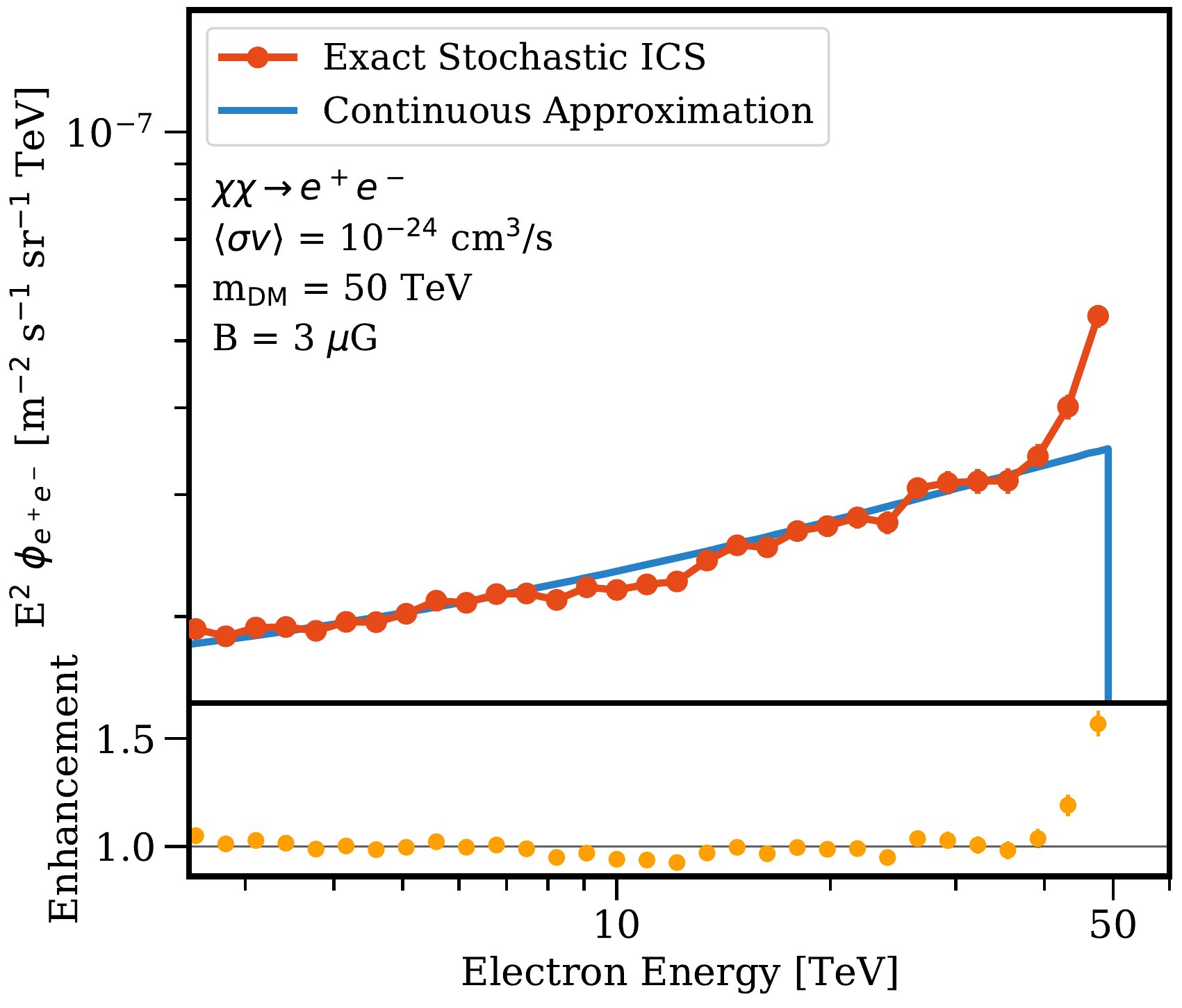} 
\end{minipage}
\hfill
\begin{minipage}[t]{0.48\textwidth}
\includegraphics[width=0.98\textwidth]{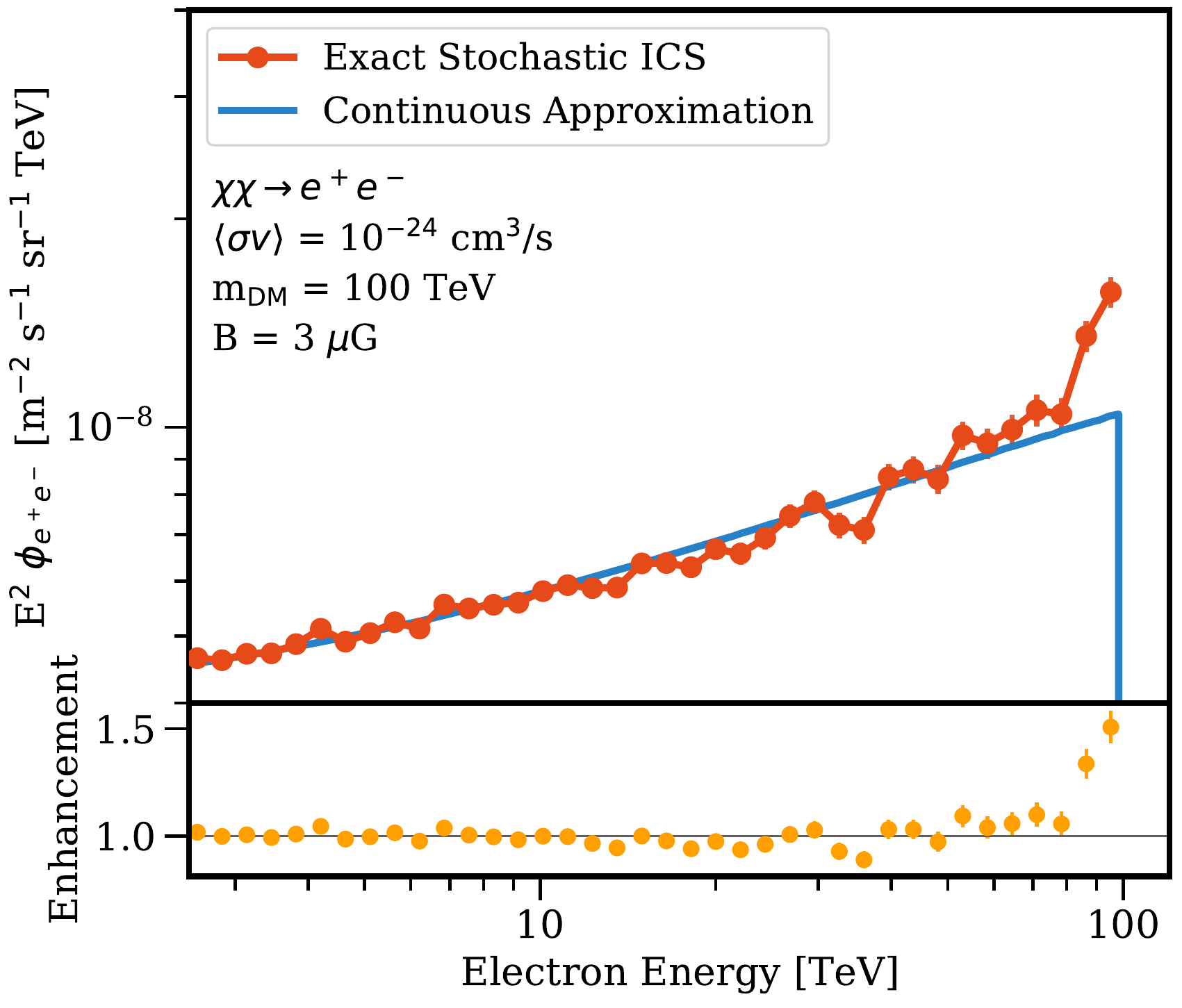} 
\end{minipage}
\vfill
\begin{minipage}[t]{0.48\textwidth}
\includegraphics[width=0.98\textwidth]{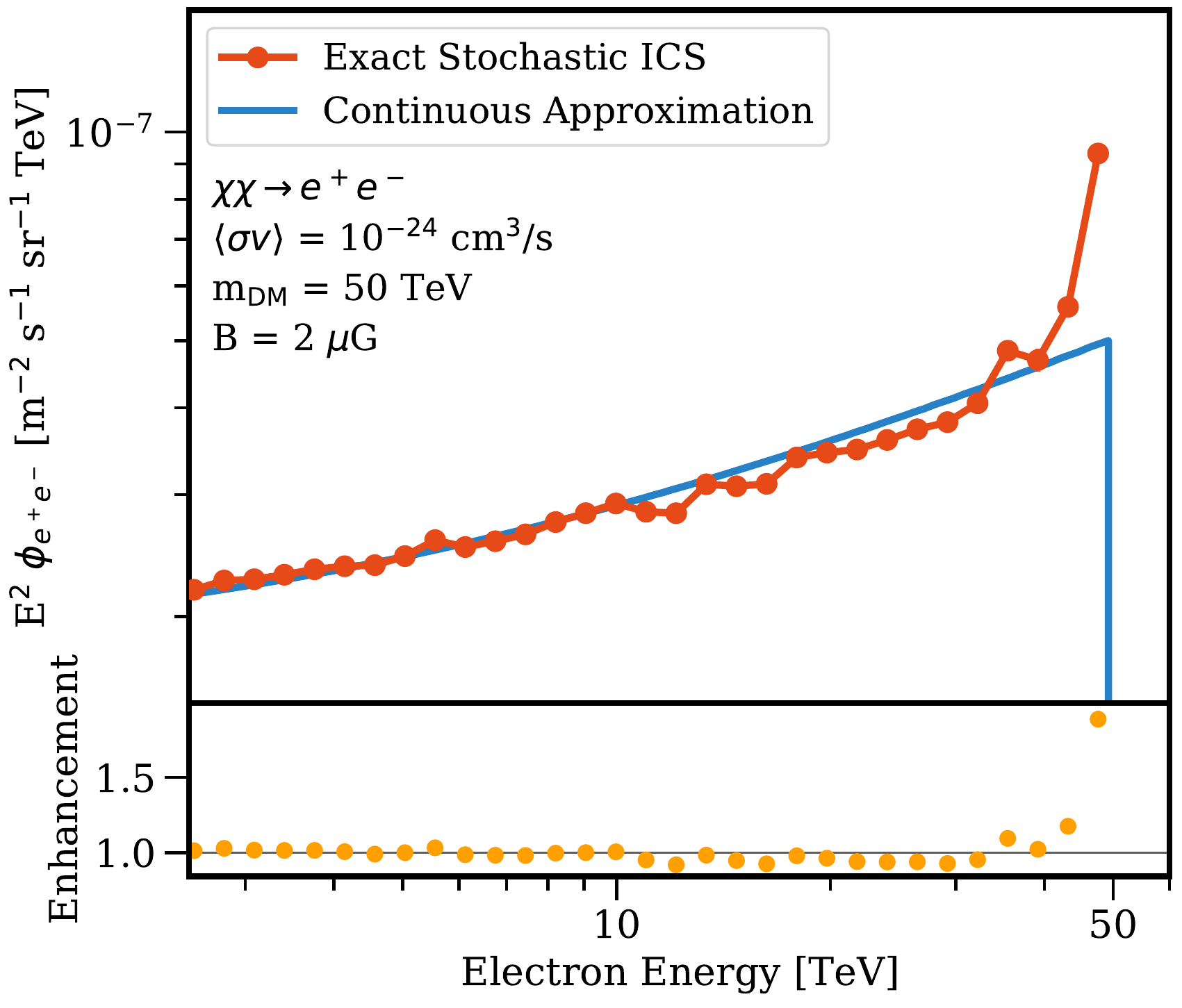} 
\end{minipage}
\hfill
\begin{minipage}[t]{0.48\textwidth}
\includegraphics[width=0.98\textwidth]{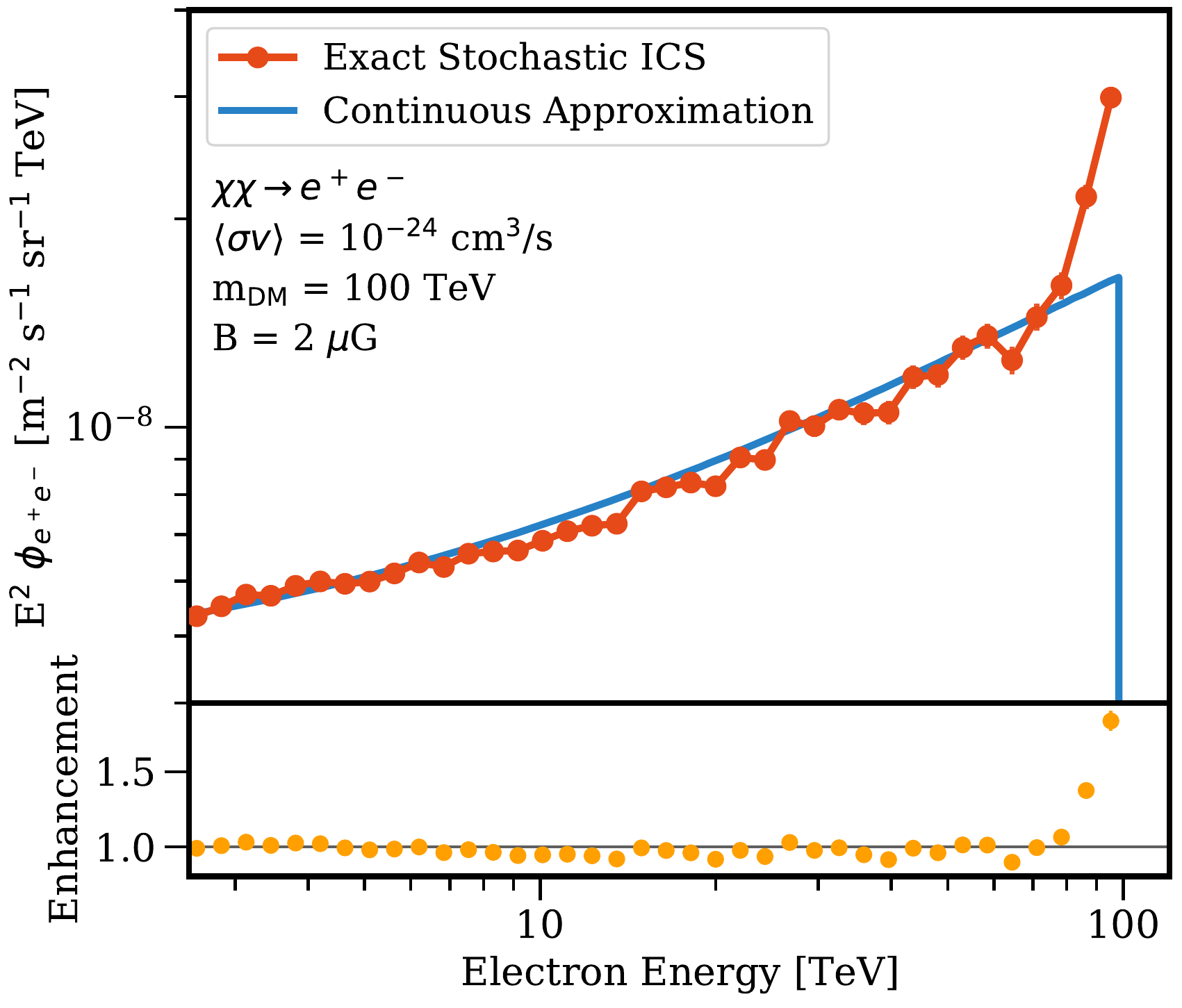} 
\end{minipage}
\vfill
\begin{minipage}[t]{0.48\textwidth}
\includegraphics[width=0.98\textwidth]{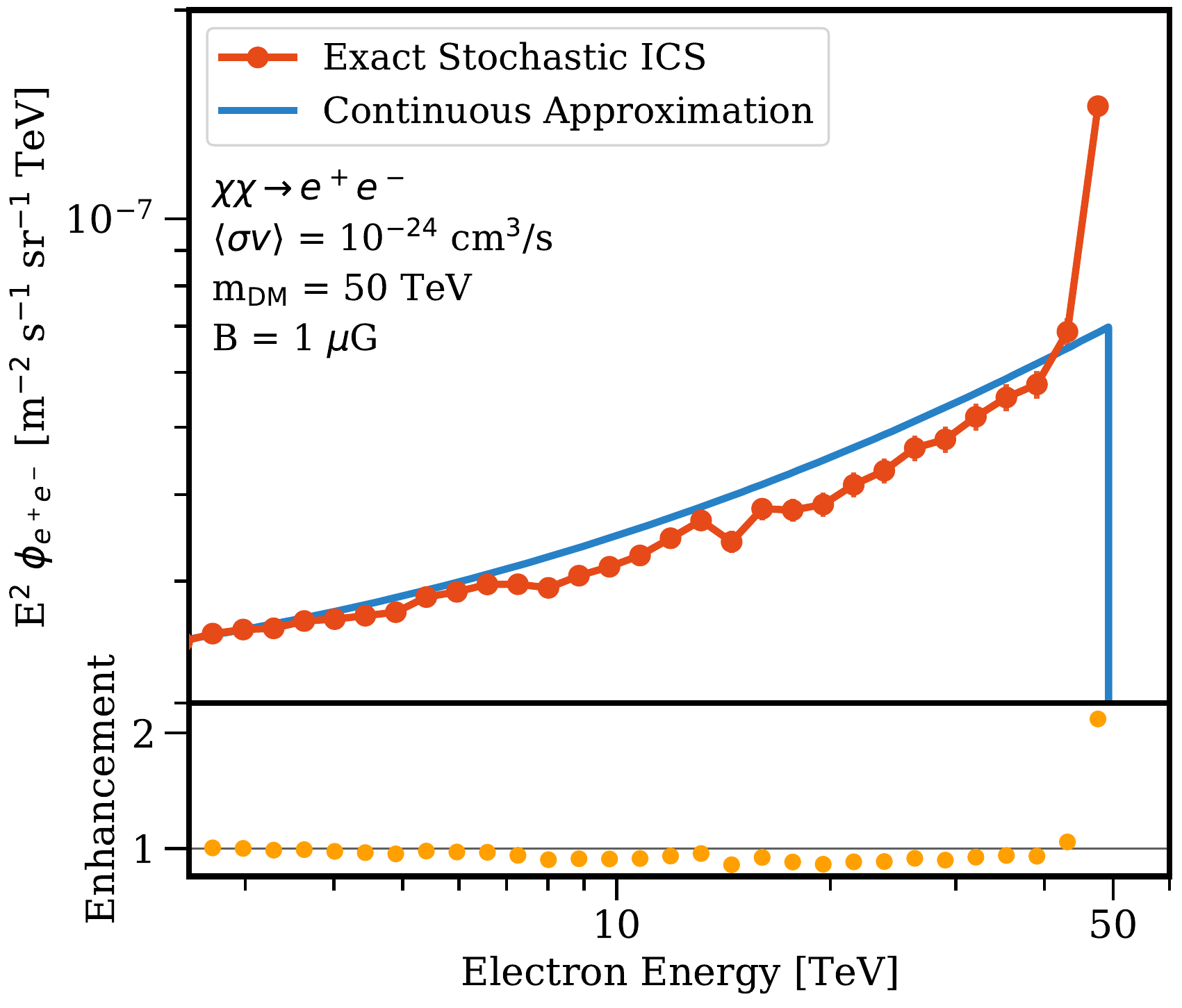} 
\end{minipage}
\hfill
\begin{minipage}[t]{0.48\textwidth}
\includegraphics[width=0.98\textwidth]{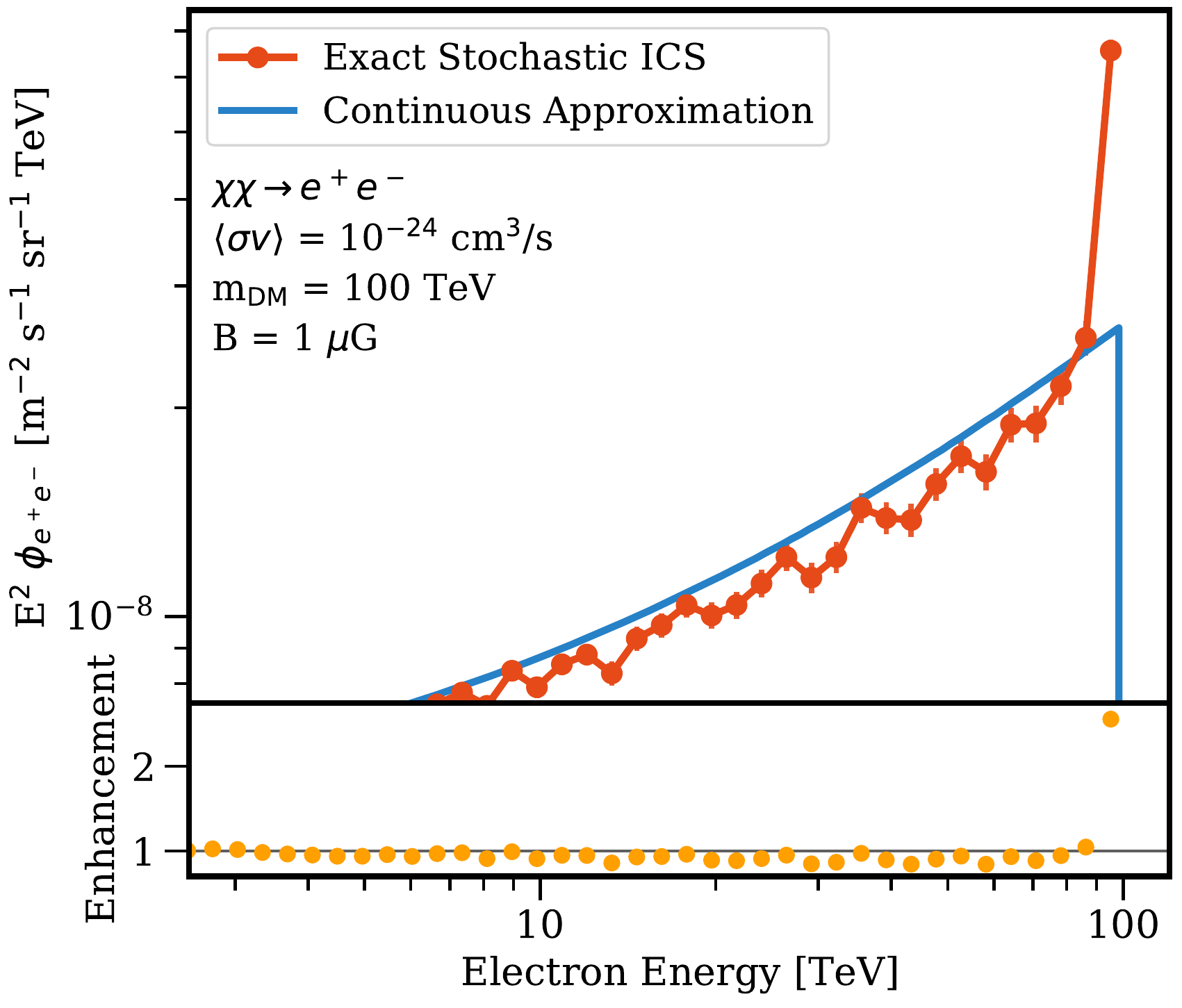} 
\end{minipage}
\caption{The expected $e^+e^-$ flux from the stochastic model (red) compared to the continuous approximation (blue), for a dark matter mass of 50~TeV (left panels) and 100~TeV (right panels) for the three different magnetic field strengths.}
\label{fig: 50 TeV 100 TeV}
\end{figure*}

\begin{figure}
\centering
\begin{minipage}[t]{0.48\textwidth}
\includegraphics[width=0.98\textwidth]{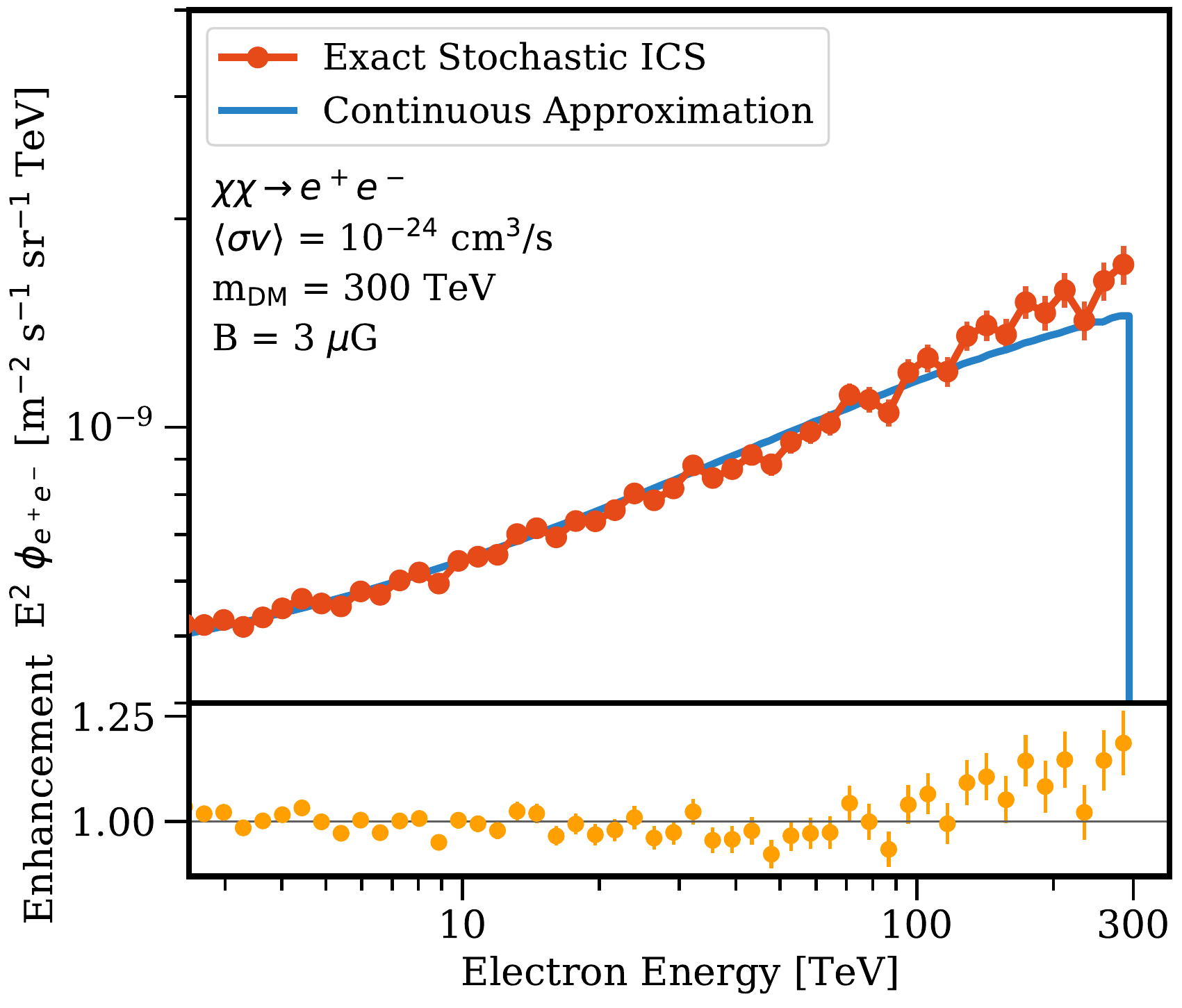} 
\end{minipage}
\hfill
\begin{minipage}[t]{0.48\textwidth}
\includegraphics[width=0.98\textwidth]{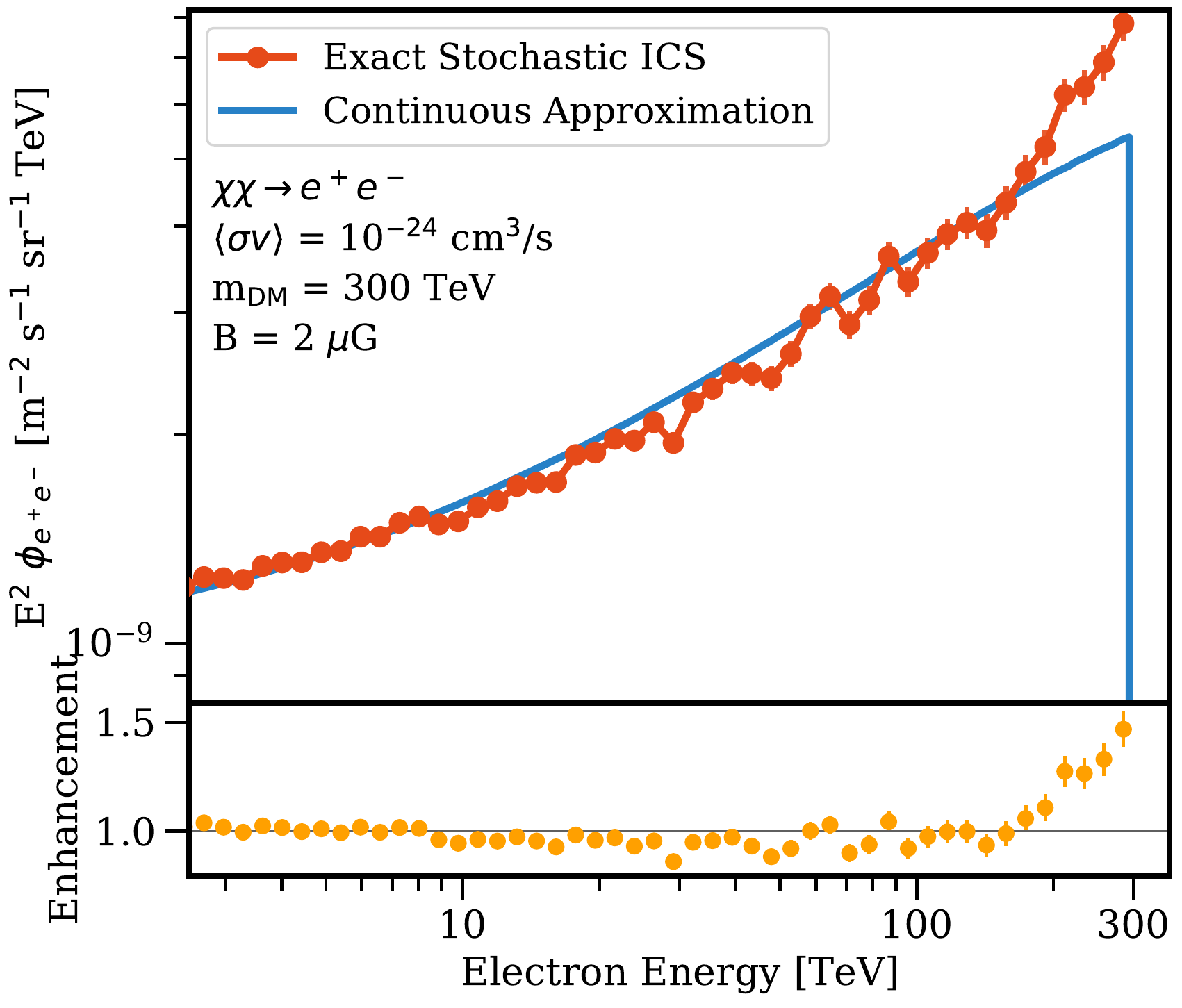} 
\end{minipage}
\hfill
\begin{minipage}[t]{0.48\textwidth}
\includegraphics[width=0.98\textwidth]{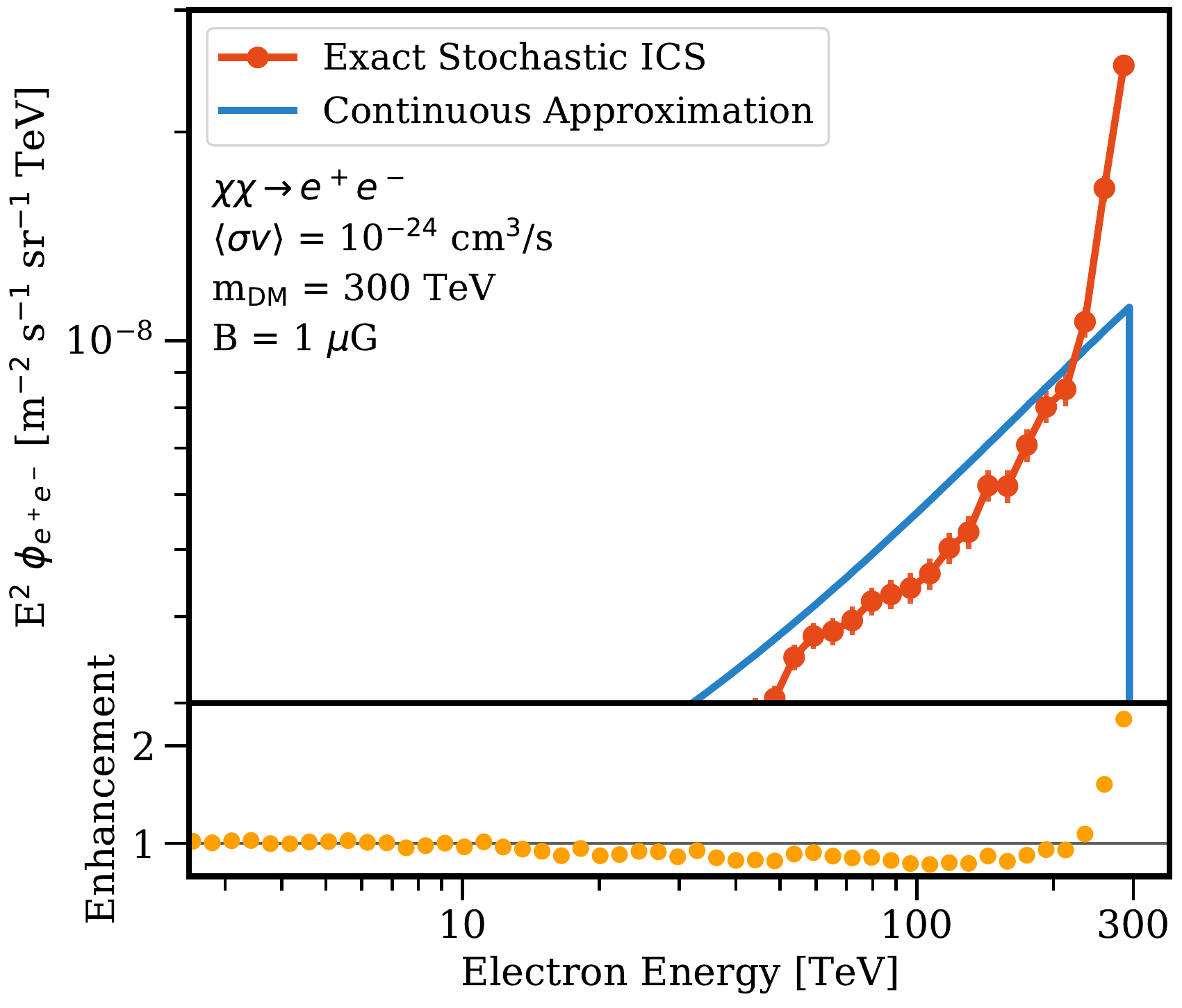} 
\end{minipage}
\caption{The expected $e^+e^-$ flux from the stochastic model (red) compared to the continuous approximation (blue), for a dark matter mass of 300~TeV for the three different magnetic field strengths.}
\label{fig: 300 TeV}
\end{figure}

In Figures~\ref{fig: 10 TeV 30 TeV} to \ref{fig: 300 TeV}, we show the $e^+e^-$ flux for dark matter annihilations into $e^+e^-$ for all our cases, which includes dark matter masses of 10, 30, 50, 100 and 300~TeV, and magnetic field strengths of 1, 2 and 3~$\mu$G. Note that for completeness, the spectra that are presented Figures~\ref{fig: fig1},  \ref{fig: 300 TeV 1 muG} and~\ref{fig: 10 TeV 3 muG} in the main text are shown here as well.

These figures show that the enhancement of the dark matter signal from the stochastic model is strongest for higher dark matter masses, because energy losses in a single interaction are larger, and smaller magnetic field strengths, because energy losses from synchrotron radiation are smaller. For example, the top right panel of Figure~\ref{fig: 50 TeV 100 TeV} shows the signal from a 100-TeV dark matter particle for a magnetic field strength of 3~$\mu$G. The stochastic model enhances the signal by a factor of $\sim$1.5, while in the bottom right panel, where the magnetic field strength is 1~$\mu$G, the enhancement is $\sim$2.6.

At energies above about 100~TeV, synchrotron energy losses start to become more efficient than ICS, and $e^+e^-$ can lose a substantial fraction of their initial energy before undergoing an ICS event. This can be seen in Figure~\ref{fig: 300 TeV} for a dark matter mass of 300~TeV, where the enhancement of the signal is slightly smaller for all three magnetic field strengths compared to the 100-TeV case, where synchrotron losses are still subdominant.

For smaller dark matter masses, e.g.~Figure~\ref{fig: 10 TeV 30 TeV} shows 10 and 30 TeV, respectively, the enhancement is also less prominent. This is because the average energy lost in the first ICS interaction is only a small fraction of the initial energy. Thus, this can be reconciled by using a higher energy resolution (see also Figure~\ref{fig: E first ICS vs E init} in the main text), which would be able to resolve the enhancement more precisely. We show this in Figures~9 and~10 in the main text for a dark matter mass of 10~TeV and magnetic field strength of 3~$\mu$G given an experiment with an energy resolution changed from 5\% to 3\% and 1\%, respectively, as well as in the Appendix in Figures~\ref{fig: 10 TeV 1 muG 0.03} and~\ref{fig: 10 TeV 1 muG 0.01} for a dark matter mass of 10~TeV and magnetic field strength of 1~$\mu$G, where the enhancement becomes a factor of $\sim$ 2 -- 3.3.

\begin{figure*}
\centering
\begin{minipage}[t]{0.48\textwidth}
\includegraphics[width=0.98\textwidth]{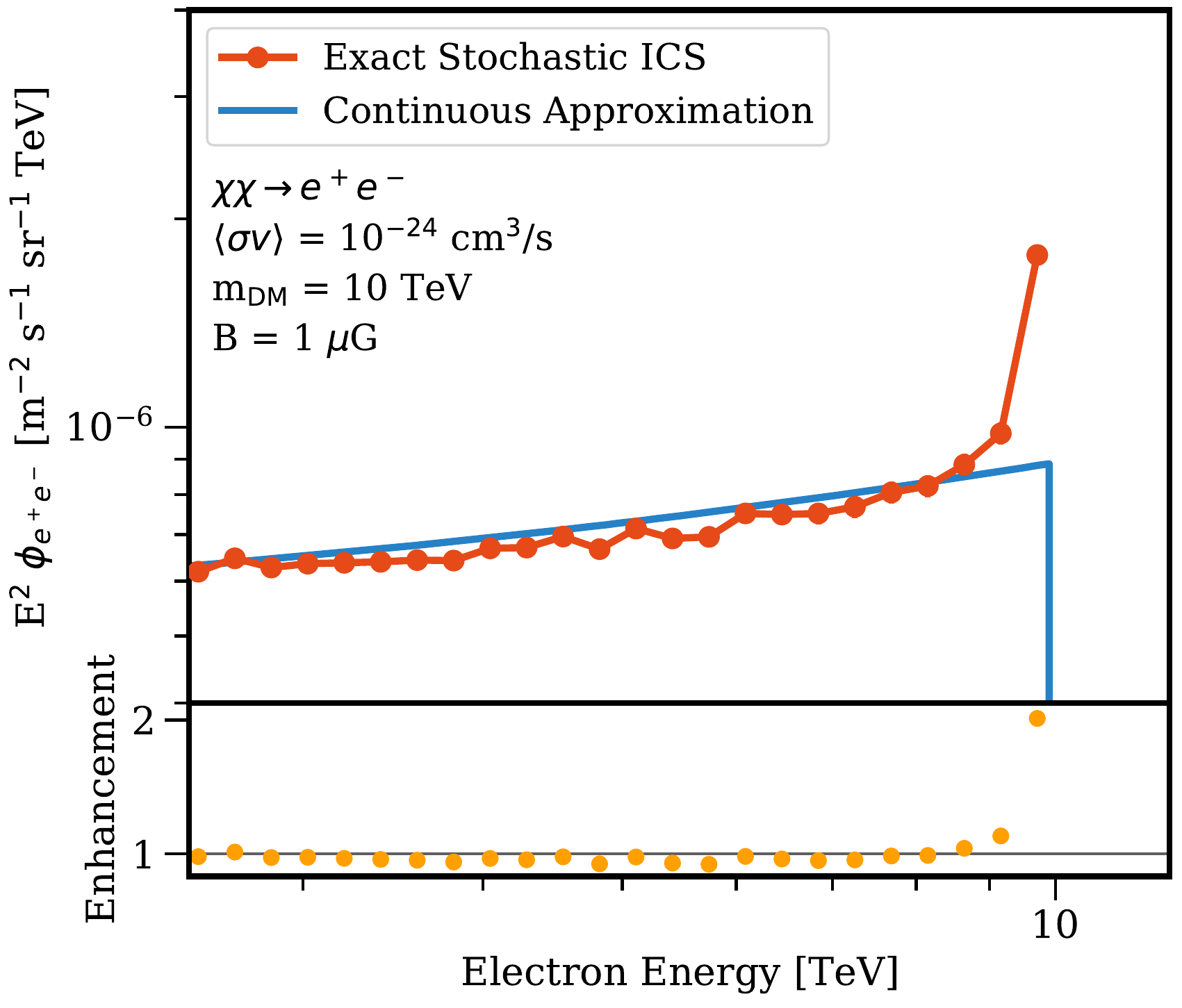} 
\caption{The expected $e^+e^-$ flux from the stochastic model (red) compared to the continuous approximation (blue), for a dark matter mass of 10~TeV and a magnetic field strength of 1~$\mu$G, similar to the bottom left panel of Figure~\ref{fig: 10 TeV 30 TeV}, but with an energy resolution of 3\% instead of 5\%. The improved resolution increases the enhancement from a factor of 1.6 to 2.1.}
\label{fig: 10 TeV 1 muG 0.03}
\end{minipage}
\hfill
\begin{minipage}[t]{0.48\textwidth}
\includegraphics[width=0.98\textwidth]{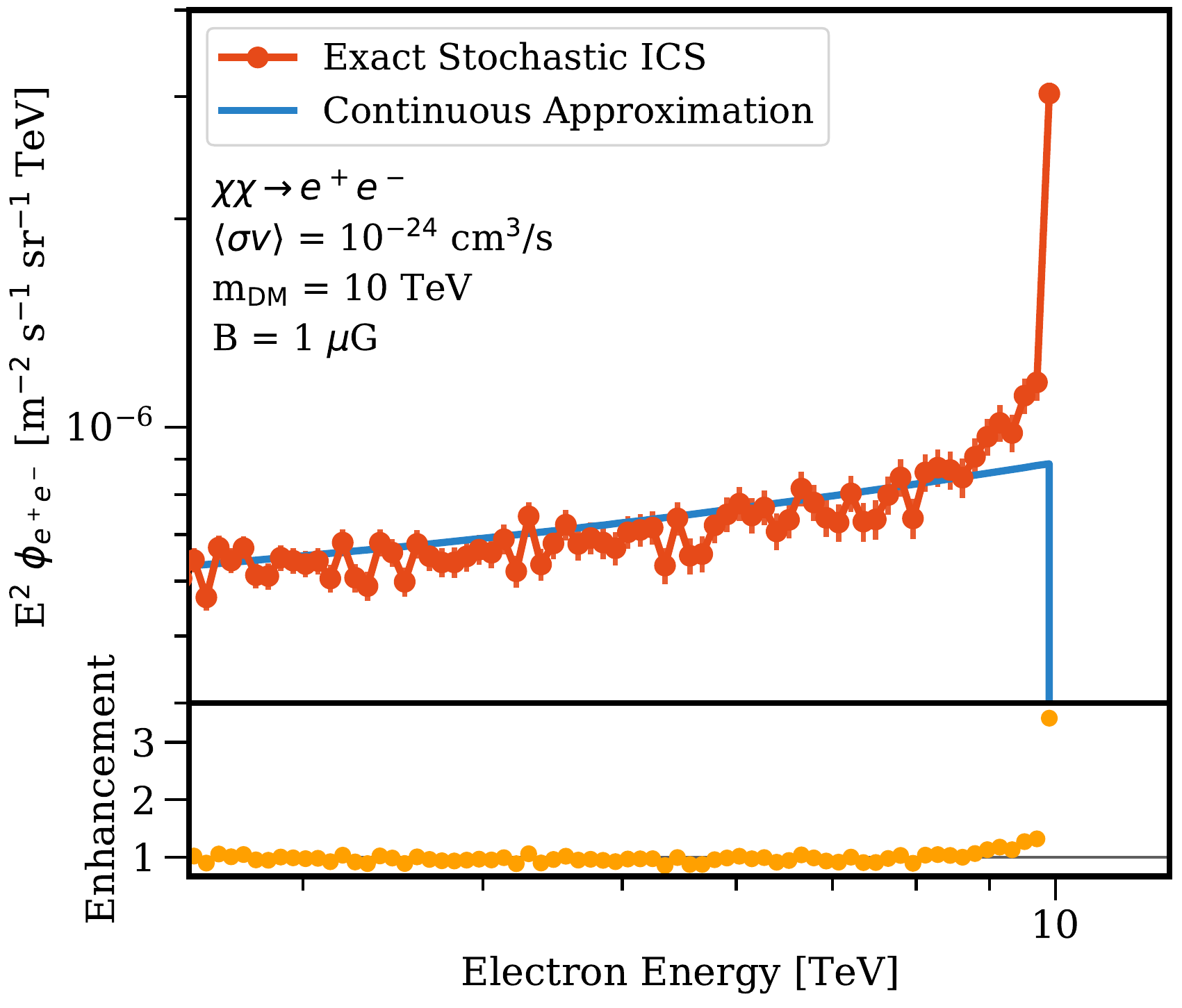} 
\caption{The expected $e^+e^-$ flux from the stochastic model (red) compared to the continuous approximation (blue), for a dark matter mass of 10~TeV and a magnetic field strength of 1~$\mu$G, similar to the bottom left panel of Figure~\ref{fig: 10 TeV 30 TeV}, but with an energy resolution of 1\% instead of 5\%. The improved resolution increases the enhancement from a factor of 1.6 to 3.3.}
\label{fig: 10 TeV 1 muG 0.01}
\end{minipage}
\end{figure*}

\end{document}